\documentclass[longauth]{aa}  

\usepackage{graphicx}
\usepackage{txfonts}
\usepackage{hyperref}
\hypersetup{
    colorlinks=true,
    linkcolor=blue,
    filecolor=magenta,      
    urlcolor=cyan,
    citecolor=blue
    }
\usepackage[dvipsnames]{xcolor}

\let\OLDthebibliography\thebibliography
\renewcommand\thebibliography[1]{
  \OLDthebibliography{#1}
  \setlength{\parskip}{0pt}
  \setlength{\itemsep}{0pt plus 0.3ex}
}
\usepackage[switch, modulo]{lineno} 

\begin{document}

   \title{Cosmological constraints from the \textit{Planck} cluster catalogue with DES shear profiles and \textit{Chandra} observations}

   \author{G.~Aymerich\inst{1,2} \fnmsep\thanks{\email{gaspard.aymerich@universite-paris-saclay.fr}} \and
        S.~Grandis\inst{3} \and
        M.~Douspis\inst{1} \and
        G.~W. Pratt\inst{2} \and
        L.~Salvati\inst{1} \and
        F.~Andrade-Santos\inst{4,5} \and 
        S.~Bocquet\inst{6} \and
        M.~Costanzi\inst{7,8,9} \and
        W.~R.~Forman\inst{4} \and 
        C.~Jones\inst{4} \and
M.~Aguena\inst{8,10} \and F.~Andrade-Oliveira\inst{11} \and D.~Bacon\inst{12} \and D.~Brooks\inst{13} \and D.~L.~Burke\inst{14,15} \and J.~Carretero\inst{16} \and L.~N.~da Costa\inst{10} \and M.~E.~da Silva Pereira\inst{17} \and T.~M.~Davis\inst{18} \and J.~De~Vicente\inst{19} \and S.~Desai\inst{20} \and H.~T.~Diehl\inst{21} \and P.~Doel\inst{13} \and S.~Everett\inst{22} \and B.~Flaugher\inst{21} \and J.~Frieman\inst{23,21,24} \and E.~Gaztanaga\inst{25,12,26} \and D.~Gruen\inst{27} \and G.~Gutierrez\inst{21} \and S.~R.~Hinton\inst{18} \and D.~L.~Hollowood\inst{28} \and K.~Honscheid\inst{29,30} \and D.~J.~James\inst{31} \and S.~Lee\inst{32} \and J.~L.~Marshall\inst{33} \and J. Mena-Fern{\'a}ndez\inst{34} \and R.~Miquel\inst{35,16} \and J.~J.~Mohr\inst{27} \and R.~L.~C.~Ogando\inst{36} \and A.~A.~Plazas~Malag\'on\inst{14,15} \and A.~Porredon\inst{19,37} \and J.~Prat\inst{23,38} \and A.~K.~Romer\inst{39} \and S.~Samuroff\inst{40,16} \and E.~Sanchez\inst{19} \and D.~Sanchez Cid\inst{19,11} \and M.~Smith\inst{41} \and E.~Suchyta\inst{42} \and M.~E.~C.~Swanson\inst{43} \and D.~L.~Tucker\inst{21} \and N.~Weaverdyck\inst{44,45} \and J.~Weller\inst{46,47} \and M.~Yamamoto\inst{48,49} 
          }

   \institute{
         Université Paris-Saclay, CNRS, Institut d'Astrophysique Spatiale, 91405, Orsay, France \and Université Paris-Saclay, Université Paris Cité, CEA, CNRS, AIM, 91191, Gif-sur-Yvette, France \and Universität Innsbruck, Institut für Astro- und Teilchenphysik, Technikerstr. 25/8, 6020 Innsbruck, Austria \and Center for Astrophysics $\vert$ Harvard \& Smithsonian, Cambridge, MA 02138, USA \and Department of Liberal Arts and Sciences, Berklee College of Music, 7 Haviland Street, Boston, MA 02215, USA \and University Observatory, Faculty of Physics, Ludwig-Maximilians-Universität, Scheinerstr. 1, 81679 Munich, Germany \and Astronomy Unit, Department of Physics, University of Trieste, Via Tiepolo 11, I-34131 Trieste, Italy \and INAF-Osservatorio Astronomico di Trieste, Via G. B. Tiepolo 11, I-34143 Trieste, Italy \and Institute for Fundamental Physics of the Universe, Via Beirut 2, 34014 Trieste, Italy \and Laborat\'orio Interinstitucional de e-Astronomia - LIneA, Av. Pastor Martin Luther King Jr, 126 Del Castilho, Nova Am\'erica Offices, Torre 3000/sala 817 CEP: 20765-000, Brazil \and Physik-Institut, University of Zürich, Winterthurerstrasse 190, CH-8057 Zürich, Switzerland \and Institute of Cosmology and Gravitation, University of Portsmouth, Portsmouth, PO1 3FX, UK \and Department of Physics \& Astronomy, University College London, Gower Street, London, WC1E 6BT, UK \and Kavli Institute for Particle Astrophysics \& Cosmology, P. O. Box 2450, Stanford University, Stanford, CA 94305, USA \and SLAC National Accelerator Laboratory, Menlo Park, CA 94025, USA \and Institut de F\'{\i}sica d'Altes Energies (IFAE), The Barcelona Institute of Science and Technology, Campus UAB, 08193 Bellaterra (Barcelona) Spain \and Hamburger Sternwarte, Universit\"{a}t Hamburg, Gojenbergsweg 112, 21029 Hamburg, Germany \and School of Mathematics and Physics, University of Queensland,  Brisbane, QLD 4072, Australia \and Centro de Investigaciones Energ\'eticas, Medioambientales y Tecnol\'ogicas (CIEMAT), Madrid, Spain \and Department of Physics, IIT Hyderabad, Kandi, Telangana 502285, India \and Fermi National Accelerator Laboratory, P. O. Box 500, Batavia, IL 60510, USA \and California Institute of Technology, 1200 East California Blvd, MC 249-17, Pasadena, CA 91125, USA \and Department of Astronomy and Astrophysics, University of Chicago, Chicago, IL 60637, USA \and Kavli Institute for Cosmological Physics, University of Chicago, Chicago, IL 60637, USA \and Institut d'Estudis Espacials de Catalunya (IEEC), 08034 Barcelona, Spain \and Institute of Space Sciences (ICE, CSIC),  Campus UAB, Carrer de Can Magrans, s/n,  08193 Barcelona, Spain \and University Observatory, LMU Faculty of Physics, Scheinerstr. 1, 81679 Munich, Germany \and Santa Cruz Institute for Particle Physics, Santa Cruz, CA 95064, USA \and Center for Cosmology and Astro-Particle Physics, The Ohio State University, Columbus, OH 43210, USA \and Department of Physics, The Ohio State University, Columbus, OH 43210, USA \and Center for Astrophysics $\vert$ Harvard \& Smithsonian, 60 Garden Street, Cambridge, MA 02138, USA \and Jet Propulsion Laboratory, California Institute of Technology, 4800 Oak Grove Dr., Pasadena, CA 91109, USA \and George P. and Cynthia Woods Mitchell Institute for Fundamental Physics and Astronomy, and Department of Physics and Astronomy, Texas A\&M University, College Station, TX 77843,  USA \and Universit\'e Grenoble Alpes, CNRS, LPSC-IN2P3, 38000 Grenoble, France \and Instituci\'o Catalana de Recerca i Estudis Avan\c{c}ats, E-08010 Barcelona, Spain \and Observat\'orio Nacional, Rua Gal. Jos\'e Cristino 77, Rio de Janeiro, RJ - 20921-400, Brazil \and Ruhr University Bochum, Faculty of Physics and Astronomy, Astronomical Institute, German Centre for Cosmological Lensing, 44780 Bochum, Germany \and Nordita, KTH Royal Institute of Technology and Stockholm University, Hannes Alfv\'ens v\"ag 12, SE-10691 Stockholm, Sweden \and Department of Physics and Astronomy, Pevensey Building, University of Sussex, Brighton, BN1 9QH, UK \and Department of Physics, Northeastern University, Boston, MA 02115, USA \and Physics Department, Lancaster University, Lancaster, LA1 4YB, UK \and Computer Science and Mathematics Division, Oak Ridge National Laboratory, Oak Ridge, TN 37831 \and Center for Astrophysical Surveys, National Center for Supercomputing Applications, 1205 West Clark St., Urbana, IL 61801, USA \and Department of Astronomy, University of California, Berkeley,  501 Campbell Hall, Berkeley, CA 94720, USA \and Lawrence Berkeley National Laboratory, 1 Cyclotron Road, Berkeley, CA 94720, USA \and Max Planck Institute for Extraterrestrial Physics, Giessenbachstrasse, 85748 Garching, Germany \and Universit\"ats-Sternwarte, Fakult\"at f\"ur Physik, Ludwig-Maximilians Universit\"at M\"unchen, Scheinerstr. 1, 81679 M\"unchen, Germany \and Department of Astrophysical Sciences, Princeton University, Peyton Hall, Princeton, NJ 08544, USA \and Department of Physics, Duke University Durham, NC 27708, USA  
            }

   \date{DES-2025-0899, FERMILAB-PUB-25-0432-PPD \\
   Received XXX; accepted YYY}

  \abstract{We present cosmological constraints from the \textit{Planck} PSZ2 cosmological cluster sample, using weak-lensing shear profiles from Dark Energy Survey (DES) data and X-ray observations from the \textit{Chandra} telescope for the mass calibration. We compute hydrostatic mass estimates for all clusters in the PSZ2 sample with a scaling relation between their integrated Sunyaev-Zeldovich signal and X-ray derived hydrostatic mass, calibrated with the \textit{Chandra} data. We introduce a method to correct these masses with a hydrostatic mass bias using shear profiles derived from wide-field galaxy surveys. We simultaneously fit the number counts of the PSZ2 sample and the mass calibration with the DES data, finding $\Omega_\text{m}=0.312^{+0.018}_{-0.024}$, $\sigma_8=0.777\pm 0.024$, $S_8\equiv \sigma_8 \sqrt{\Omega_\text{m} / 0.3}=0.791^{+0.023}_{-0.021}$, and $(1-b)=0.844^{+0.055}_{-0.062}$ for our baseline analysis when combined with BAO data. When considering a hydrostatic mass bias evolving with mass, we find $\Omega_\text{m}=0.353^{+0.025}_{-0.031}$, $\sigma_8=0.751\pm 0.023$, and $S_8=0.814^{+0.019}_{-0.020}$. We verify the robustness of our results by exploring a variety of analysis settings, with a particular focus on the definition of the halo centre used for the extraction of shear profiles. We compare our results with a number of other cosmological analyses, in particular two recent analyses of cluster samples obtained from SPT and eROSITA data that share the same mass calibration data set. We find that our results are in overall agreement with most late-time probes, in very mild tension with CMB results (1.6$\sigma$), and in significant tension with results from eROSITA clusters (2.9$\sigma$). We confirm that our mass calibration is consistent with the eROSITA analysis by comparing masses for clusters present in both \textit{Planck} and eROSITA samples, eliminating it as a potential cause of tension.}


    \titlerunning{Cosmological constraints from the PSZ2 catalogue with DES calibration}
    \authorrunning{Aymerich et al.}
   \maketitle
%

\section{Introduction}
\label{intro}
The standard $\Lambda$CDM model is the most widely accepted theoretical framework to describe the evolution of our Universe. It is the most minimal model that is able to describe all of our observations, containing only six free parameters and adopting a very simple modelling of dark matter and dark energy. Over the years, several experiments have put increasingly tight constraints on the value of the $\Lambda$CDM parameters, by comparing different probes with theoretical predictions. Some common probes include the primary anisotropies in the cosmic microwave background (CMB) \citep[see e.g.][]{planckcollaborationvi_planck_2020, balkenhol_measurement_2023, louis_atacama_2025}, the large scale structures of the Universe observed via weak-lensing \citep[see e.g.][]{heymans_kids1000_2021, amon_dark_2022, li_hyper_2023} and surveys of galaxy clusters \citep[see e.g.][]{planckcollaborationxxiv_planck_2016, bocquet_spt_2024a, aymerich_cosmological_2024, ghirardini_srg_2024}. While the $\Lambda$CDM model is very successful at predicting the observations individually, some tensions arise when comparing the preferred values of the model's free parameters derived by different types of experiments. One of these tensions concerns the value of the $S_8$ parameter, defined as $S_8 \equiv \sigma_8 \sqrt{\Omega_\text{m} / 0.3}$, to combine the often degenerated matter density $\Omega_\text{m}$ and amplitude of the mass fluctuations on scales of 8 Mpc/h $\sigma_8$ parameters in a single number. Most observations of the late-time Universe, like galaxy cluster surveys \citep[e.g.][]{planckcollaborationxxiv_planck_2016, costanzi_cosmological_2021, aymerich_cosmological_2024, bocquet_spt_2024a} and especially large-scale structure surveys \citep[e.g.][]{amon_dark_2022, li_hyper_2023, wright_kidslegacy_2025}, prefer a lower $S_8$ value by 2 to 3$\sigma$ compared to the one derived from CMB primary anisotropies observations like \cite{planckcollaborationvi_planck_2020}. The recent publication of the first cosmological results from the eROSITA galaxy cluster sample \citep{ghirardini_srg_2024} was a major update to the status of the $S_8$ tension. This analysis derived the tightest $S_8$ constraints from a late-time Universe probe to date, and found $S_8 = 0.86\pm 0.01$. This value is larger than the $S_8 = 0.832\pm 0.013$ constraint from \cite{planckcollaborationvi_planck_2020}, in tension with most galaxy cluster or large scale structure surveys. The latest analysis of clusters detected in the Dark Energy Survey (DES) also yielded a high $S_8$ value compared to most previous late-time Universe probes, finding $S_8 = 0.864\pm 0.035$, or $S_8 = 0.811^{+0.022}_{-0.020}$ when combined with 3$\times$2pt \citep{descollaboration_dark_2025}.

Galaxy clusters are the largest virialized objects in the Universe and can be used as tracers of structure formation. Their abundance as a function of mass and redshift is a cosmological probe, particularly sensitive to the value of $\Omega_\text{m}$ and $\sigma_8$. They are multi-component objects, containing dark matter, galaxies, and hot gas and can thus be observed at different wavelengths. The hot gas emits through thermal bremsstrahlung in the X-ray domain, and can also be detected through its interaction with the CMB photons, named the thermal Sunyaev-Zeldovich (SZ) effect \citep{sunyaev_observations_1972}. In the optical domain, the member galaxies can be observed, and the gravitational potential well of the cluster can also be detected by observing the gravitational lensing of background galaxies. 

To conduct a cosmological analysis of galaxy cluster number counts, three ingredients are needed: a theoretical prediction of cluster abundance as a function of cosmology, a catalogue of galaxy clusters with a well-understood selection function, and a link between the observables of galaxy clusters and their total mass at a population level. This last step, commonly referred to as the mass-observable relation or mass calibration, is currently the limiting factor of galaxy cluster analyses, with the error budget being dominated by the systematics of the mass calibration and not the statistics of the cluster catalogue in most studies \citep[see][for a review of galaxy cluster mass calibration]{pratt_galaxy_2019}. The characteristic signature of the hot gas, or intra-cluster medium (ICM), is generally used to construct the galaxy cluster catalogue, either in the microwave \citep[see e.g.][]{planckcollaborationxxvii_planck_2016, bleem_galaxy_2015, bleem_sptpol_2020, bleem_galaxy_2024, hilton_atacama_2021, klein_sptsz_2024} or X-ray sky \citep[see e.g.][]{pacaud_xxl_2018, bulbul_srg_2024}, but some studies also used optical data to detect clusters \citep[see e.g.][]{rozo_redmapper_2015, mcclintock_dark_2019, maturi_amico_2019, chiu_weaklensing_2024}. However, the baryonic matter, in the form of either member galaxies or the ICM, represents only $\sim$20\% of the total mass of galaxy clusters, and strong assumptions, like hydrostatic equilibrium, need to be made to compute masses directly from observations of the baryonic matter. Since weak gravitational lensing (WL) is caused by the gravitational potential of a cluster, it provides a measurement of the total mass of the cluster, including the dark matter, resulting in mass estimates that are independent of assumptions about the dynamical state of the system. This leads most cosmological analyses of galaxy clusters to include weak-lensing data in the mass calibration process \citep[see e.g.][for one of the first applications of WL to cluster cosmology]{mantz_cosmology_2015}. This was previously only possible through pointed observations of individual clusters \citep[like the WL data used in][for example]{planckcollaborationxxiv_planck_2016}, which have the drawback of requiring a significant amount of observational time, especially with the constantly increasing size of cluster samples. Recent advances in measuring, calibrating, and interpreting the weak gravitational lensing signal measured in the DES shape catalogues around ICM-selected clusters now enable the use of wide-field galaxy survey data for cluster mass calibration \citep[see][]{grandis_calibration_2021, bocquet_spt_2024, grandis_srg_2024}. Notably, this mass calibration has been adopted in two of the latest ICM-selected galaxy cluster analyses: \cite{bocquet_spt_2024a}, using an SZ-detected cluster catalogue from South Pole Telescope (SPT) data, and \cite{ghirardini_srg_2024}, using an X-ray detected cluster sample from eROSITA data. Interestingly, the two studies yielded different final cosmological constraints: \cite{bocquet_spt_2024a} found results consistent with most late-time probes, while \cite{ghirardini_srg_2024} reported a higher $S_8$ value. Applying this new mass calibration procedure to a third cluster catalogue may help clarify these differences and improve our understanding of their cosmological implications.

In this work, we study the abundance of galaxy clusters detected through their SZ signature in the \textit{Planck} sky. While the \textit{Planck} sample is not as large as some of the latest catalogues, it is a very well understood sample that has been studied in depth in the literature \citep[see e.g.][]{planckcollaborationxxvii_planck_2016, gallo_characterising_2024, saxena_chexmate_2025}. However, significant uncertainties remain in the mass calibration of the \textit{Planck} cluster sample, and the sample size is currently not a limiting factor for its constraining power. This work presents the constraints on cosmological parameters obtained from the \textit{Planck} cluster sample, using X-ray follow-up observations and tangential shear profiles derived from DES data for the mass calibration. We study the problem of centre definition for \textit{Planck} detected clusters and its impact on the final constraints, and also provide a forecast of future potential constraining power of the \textit{Planck} cluster sample with Stage IV WL lensing data using mock data. 

We build on the analysis presented in \citet{aymerich_cosmological_2024}, which improved on the original \citet{planckcollaborationxxiv_planck_2016} study by updating the mass calibration with newer X-ray and WL data sets, keeping the overall pipeline unchanged where possible. The first step, that is shared with this work, was the calibration of a scaling relation between X-ray derived hydrostatic masses and signal in the \textit{Planck} sky with the X-ray calibration subsample. This allows for the derivation of a hydrostatic mass for every cluster in the \textit{Planck} sample. The second step was the calibration of a hydrostatic mass bias, by comparing the hydrostatic masses previously obtained with individual masses derived from pointed WL observations. Finally, the cosmology sampling of the number counts was done, using the scaling relation and hydrostatic mass bias as priors. In this work, we change the second step and combine it with the third one. Instead of using independently derived WL masses, we calibrate the hydrostatic mass bias with tangential shear profiles derived from DES data. We also perform the bias calibration simultaneously with the cosmological sampling, allowing us to properly compute distances in the WL calibration, taking the cosmology into account. The interest is two-fold. First, it allows for a direct comparison of our results with the SPT and eROSITA cluster number counts analyses by using the same mass calibration procedure. Secondly, it prepares the \textit{Planck} cluster catalogue for the upcoming Stage IV WL datasets that this analysis can be easily adapted to. While the hydrostatic mass bias is not used in some recent cluster studies, including \citet{bocquet_spt_2024a} and \citet{ghirardini_srg_2024}, we choose to include it in our analysis as it has not yet been constrained with wide-field WL data despite being an active subject of research, in both simulations \citep[see e.g.][]{gianfagna_exploring_2021, lebeau_mass_2024, kay_relativistic_2024} and observations \citep[see e.g.][]{wicker_constraining_2023, sereno_chexmate_2025, xrismcollaboration_constraining_2025}. Because it is the most common form, we adopt a fixed mass bias in our baseline analysis, but also derive cosmological constraints with a mass bias varying with mass and/or redshift.

The paper is structured as follows: we first present in Sect.~\ref{data} the different data sets used in this work. In Sect.~\ref{hydro_masses}, we present the computation of hydrostatic masses for \textit{Planck} using X-ray data. We then introduce the two parts of our likelihood: in Sect.~\ref{bias}, the first part that uses the WL data to calibrate a hydrostatic mass bias, and in Sect.~\ref{cosmo}, the second part that fits the observed number counts. In Sect.~\ref{constraints}, we present the constraints resulting from the sampling of the total likelihood, and discuss them and present forecasted future constraining power in Sect.~\ref{discussion}. We finally summarise our findings and conclude in Sect.~\ref{conclusion}.

\section{Data}
\label{data}
\subsection{PSZ2 cosmological sample}
\label{SZ_catalogue}
The \textit{Planck} PSZ2 cosmological sample, provided by the second data release of the \textit{Planck} collaboration \citep{planckcollaborationxxvii_planck_2016}, is a SZ-selected cluster catalogue, compiled from detections in the \textit{Planck} High-Frequency Instrument data. It is composed of all the galaxy clusters detected by a multi-frequency matched filter algorithm \citep[MMF;][]{melin_catalog_2006} with a signal-to-noise ratio (SNR) greater than 6 in the \textit{Planck} maps. Since purity was the main priority for the definition of this cluster catalogue, a conservative mask of the regions contaminated by Galactic emission was used, discarding 35\% of the total sky area. The PSZ2 cosmological sample contains 439 clusters and was the catalogue chosen to constrain cosmological parameters with cluster number counts in \cite{planckcollaborationxxiv_planck_2016}.

\subsection{X-ray data: \textit{Chandra} observation of \textit{Planck} ESZ sample}
\label{Chandra_Planck}
The \textit{Planck} Early Sunyaev-Zeldovich (ESZ) sample is the first cluster catalogue compiled from \textit{Planck} detections in \cite{planckcollaborationvii_planck_2011}. It consists of all the detections with SNR above 6 in the \textit{Planck} early results maps and contains 189 clusters. In this work, we use data from the \textit{Chandra-Planck} Legacy Program for Massive Clusters of Galaxies\footnote{\url{https://hea-www.cfa.harvard.edu/CHANDRA_PLANCK_CLUSTERS/}}, a full follow-up observational program with the \textit{Chandra} X-ray observatory of the 163 clusters with redshift $z<$0.35 from the \textit{Planck} ESZ catalogue. The exposure times were chosen to obtain a minimum of 10\,000 source counts per cluster. The X-ray data reduction process is described in \cite{andrade-santos_chandra_2021} \citep[see also][]{vikhlinin_chandra_2005}, and we refer the reader to these papers for more details. 

\subsection{Weak-lensing data: the DES survey}
\label{weak_lensing_des}

The DES is an approximately 5,000~deg$^2$ photometric survey in the optical bands $grizY$, carried out at the 4m Blanco telescope at the Cerro Tololo Inter-American Observatory (CTIO), Chile, with the Dark Energy Camera \citep[DECam][]{flaugher_dark_2015}.  In this analysis, we use data from the first three years of observations (DES~Y3), covering the full survey footprint.

\begin{figure}[]
    \centering
    \includegraphics[width=\columnwidth]{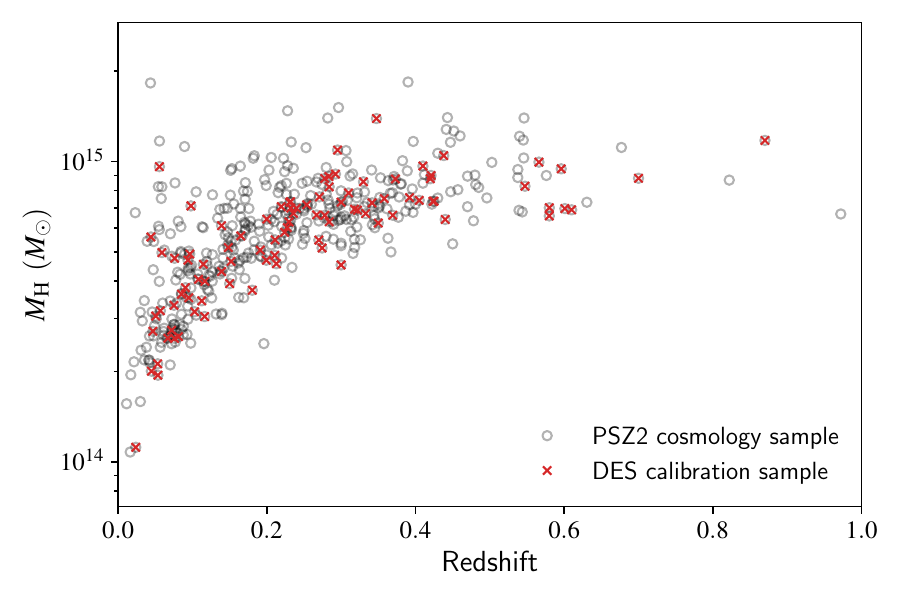}
    \caption{Mass and redshift distribution of the PSZ2 and DES calibration samples. The masses used for the plot are the hydrostatic masses derived in Sect.~\ref{hydro_mass_derivation}.}
    \label{fig:m_z}
\end{figure}

The DES~Y3 shape catalog \citep{gatti_dark_2021} is constructed from the $r,i,z$-bands using the \textsc{Metacalibration}
algorithm \citep{huff_metacalibration_2017, sheldon_practical_2017}.
Other DES~Y3 publications report further details about the photometric 
dataset \citep{sevilla-noarbe_dark_2021}, the Point-Spread Function (PSF) modelling \citep{jarvis_dark_2021}, and image simulations \citep{maccrann_dark_2022}. We refer the reader to these works for more information.
Covering an area of 4,143~square degrees, the DES~Y3 shear catalogue comprises approximately 100~million galaxies after applying all source selection cuts. Its effective source density is 5--6~galaxies arcmin$^{-2}$. 93 clusters from the PSZ2 cosmological sample are contained within the DES footprint. Fig.\ref{fig:m_z} shows the mass and redshift distribution of the PSZ2 cosmological sample, and highlights the clusters with available DES data, hereafter DES calibration sample. Because the only additional selection is the position on the sky, the DES calibration sample is expected to be representative of the full PSZ2 cosmological sample.

Our analysis uses the same selection of lensing source galaxies in tomographic bins as the DES cosmic shear analysis \citep{amon_dark_2022, secco22}. This selection is defined in \citet{myles_dark_2021}, which estimates the source redshifts with Self-Organizing Maps.
The final calibration accounts for the (potentially correlated) systematic uncertainties in source redshifts and shear measurements, as determined in image simulations \citep{maccrann_dark_2022}.

We will also use \textsc{dnf} \citep{devicente_dnf_2016} and \textsc{bpz} \citep{benitez_bayesian_2000} source redshift measurements to constrain the amount of cluster member contamination in our source sample (see Appendix~\ref{cluster_member_contamination}). While \textsc{dnf} uses a nearest-neighbour interpolation on the colour-magnitude space of spectroscopic reference samples, \textsc{bpz} is based on Bayesian fitting using galaxy templates. As such, \textsc{dnf} is more data-driven and avoids the biases that occur from the template choice, while \textsc{bpz} is more robust against the incompleteness of the spectroscopic reference samples \citep[see][Section 6.3, for more details]{sevilla-noarbe_dark_2021}.  

\subsection{Cluster centre definition}
\label{centers}
Knowledge of the cluster centres is required for proper analysis of the weak-lensing data. However, this is not trivial in the case of \textit{Planck}-detected clusters, as the PSF of the \textit{Planck} telescope is quite large at $\sim7$ arcmin. For reference, this corresponds to the angular size $\theta_{500}$ of a cluster of mass $M = 5 \cdot 10 ^{14} M_\odot$ at redshift $z=0.15$, meaning that many of the PSZ2 clusters are smaller than the PSF.

Assuming that the gas distribution accurately maps the underlying gravitational potential, we expect the angular distances between SZ-derived centres and actual halo centres to be distributed following a Rayleigh distribution of scale parameter $\text{PSF}/\text{SNR}$. In the case of the PSZ2 cosmological sample, given the \textit{Planck} PSF and average SNR of detections, this means that an average miscentring of $\sim1$ arcmin is expected. 

Photometric DES data can also be used to derive halo centres, by finding an optical counterpart to the SZ-detected cluster. This was done in an automated way with the \texttt{MCMF} algorithm for all \textit{Planck}-detected cluster candidates with SNR above 3 using DES data \citep{hernandez-lang_pszmcmf_2023} . This algorithm searches for red-sequence galaxy over-densities along the line-of-sight of a cluster candidate. Galaxies are weighted using both a radial and a colour filter, giving more weight to galaxies closer to the centre of the line-of-sight and to galaxies with photometric data fitting the red-sequence model. A limitation of this automated method when applied to \textit{Planck}-detected clusters is that the search area has to be large given the low precision of the SZ detections' pointing \citep[a mean aperture of $\sim 16$ arcmin was used in][]{hernandez-lang_pszmcmf_2023}, which can lead to mismatches.

Given the rather small size of the DES calibration sample, it is possible to associate a brightest cluster galaxy (BCG) to every cluster in the sample by visual inspection. The BCG has been shown to be a good tracer of the deepest point in the cluster gravitational potential \citep{lin_kband_2004} . Our BCG finding was done independently by two of the authors, with a third author making the final decision when a disagreement arose.

The observations at X-ray wavelengths can also be used to derive cluster centre position. The \textit{Chandra} X-ray observations allow for a very precise and reliable reconstruction of halo centres, given the very small PSF ($\sim0.5$ arcsec) and high exposure times. X-ray observations are not available for the full DES calibration sample, and so they cannot be directly used for the analysis, but, given their quality, they can serve as a reference position to evaluate the other centre detection methods. 

Using the \textit{Chandra} data available for 29 clusters in the DES calibration as reference centres, we fit the mis-centring distribution of the 3 centre detection methods (SZ, \texttt{MCMF} and BCG). In the case of the SZ centres, we model the angular mis-centring distribution with a single Rayleigh distribution with parameter $\sigma_1$:
\begin{equation}
  \label{eq:SZ_centre}
  P(\theta_\text{sep} \vert \sigma_1)=\frac{\theta_\text{sep}}{\sigma_1^2}\text{exp}\left( -\frac{\theta_\text{sep}^2}{2\sigma_1^2}\right).
\end{equation}
We choose to work with angular separation in the case of SZ centres as the mis-centring is expected to mainly be caused by the \textit{Planck} beam.
For the two optical centres (\texttt{MCMF} and BCG), we model the physical distance separation distribution with a mixture of two Rayleigh distributions:
\begin{multline}
  \label{eq:optical_centre}
  P(R_\text{sep}\vert \sigma_{1},\sigma_{2},f)=\\
  \frac{1}{1+f}\frac{R_\text{sep}}{\sigma_{1}^2}\text{exp}\left( -\frac{R_\text{sep}^2}{2\sigma_{1}^2}\right)+
  \frac{f}{1+f}\frac{R_\text{sep}}{\sigma_{2}^2}\text{exp}\left( -\frac{R_\text{sep}^2}{2\sigma_{2}^2}\right).
\end{multline}
We use physical distances for the optical centres as the mis-centring is expected to come from the physical separation between the BCG and brightest X-ray point (i.e. highest gas density). We also include a second Rayleigh distribution in the case of optical centres to account for potential misidentification of the BCG, and cases where the BCG is not located near the halo centre \citep[$\sim$20\% of BCG according to][]{lin_kband_2004}.

We sample the previously defined likelihoods with \texttt{emcee} \citep{foreman-mackey_emcee_2013} to fit the parameters, assuming flat priors on all parameters. Table~\ref{table:centre} reports the mean and standard deviations for all the performed fits. We find the expected mean mis-centring of $\sim$1 arcmin for the SZ centres, and find that visual BCG identification outperforms the automatic detection with MCMF, with a narrower distribution and fewer extreme outliers. For reference, at the median redshift of the sample $z=0.25$, the $\sim$1 arcmin mean SZ mis-centring corresponds to a physical separation of $\sim$240 kpc.
\begin{table}
  \caption{Best-fit mis-centring distribution parameters as defined in Eq.~\ref{eq:SZ_centre} and \ref{eq:optical_centre}.}             
  \label{table:centre}      
  \centering                          
  \begin{tabular}{c c c c}        
  \hline\hline                 
  Centre & SZ & BCG & MCMF \\    
  \hline                        
  $\sigma_1$ & $54.9\pm5.4$ arcsec & $21\pm2$ kpc & $60\pm8$ kpc\\
  $\sigma_2$ & -- & $207\pm39$ kpc & $647\pm 154$ kpc\\
  $f$ & -- & $0.54\pm0.19$ & $0.38\pm0.17$\\
     
  \hline                                   
  \end{tabular}
\end{table}

\section{$Y_{\text{SZ}}-M_\text{H}$ scaling relation and hydrostatic masses}
\label{hydro_masses}
In this work, we make use of the hydrostatic masses derived from the $Y_{\text{SZ}}-M_\text{H}$ scaling relation calibrated in \cite{aymerich_cosmological_2024}. In this section, we summarize the important steps of the mass derivation, and refer the interested reader to the original paper, where the full process is described in detail. 
\subsection{X-ray data reduction and mass computation}
For the 146 clusters of the \textit{Planck} ESZ catalogue that were observed with the \textit{Chandra} telescope, gas density and temperature can be inferred from the X-ray data. Assuming hydrostatic equilibrium, this can then be used to compute the total cluster mass. In this paragraph, we briefly describe the procedure, initially presented in \cite{andrade-santos_chandra_2021}.

Spectra and surface brightness are extracted in concentric annuli around the peak of the X-ray emission. An absorbed single-temperature thermal model is fitted to the spectrum in each bin, and the projected temperature profile thus obtained is fitted by projecting a 3D temperature model. Using this temperature model, the surface brightness can be converted to emission measure, which is then fitted with the gas density model from \cite{vikhlinin_chandra_2006}.

In \cite{vikhlinin_chandra_2009}, a scaling relation was calibrated between $Y_{\text{X}}$ and $M_\text{H}$, the mass computed assuming hydrostatic equilibrium (i.e. supposing that the gravitational potential perfectly compensates the gas pressure). $Y_{\text{X}}$, a mass proxy initially suggested in \cite{kravtsov_new_2006}, is defined as $Y_{\text{X}}=M_\text{gas} T_\text{X}^\text{exc}$, where $M_\text{gas}$ is the gas mass inside $R_{500}$, and $T_\text{X}^\text{exc}$ is the core-excised temperature obtained by fitting a single temperature model to the spectral data in the $0.15 R_{500}$ to $R_{500}$ region. This scaling relation is used to compute the mass of every cluster in the \textit{Chandra-Planck} sample.

\subsection{SZ signal extraction}
\cite{planckcollaborationxi_planck_2011} showed that, given the size of the \textit{Planck} beam (around 7 arcmin), using an external prior on a cluster size and position can lead to more robust SZ flux extraction. We thus use the X-ray derived positions and sizes as prior information when running a matched multi-frequency filter algorithm \citep{herranz_scaleadaptive_2002} to extract the SZ signal $Y_{\text{SZ}}$ from the \textit{Planck} data. We use the \textit{Planck} DR2 maps, even though newer maps are available as this work uses the PSZ2 cosmological sample obtained on DR2 maps.

The SZ frequency spectrum and the universal pressure profile from \cite{arnaud_universal_2010}, scaled to an aperture $\theta_{500}$ corresponding to $R_{500}^{\text{X-ray}}$ are convolved with the instrumental beam to serve as a template. The MMF algorithm then extracts the $Y_{\text{SZ}}$ signal by convolving the six \textit{Planck} HFI maps (100, 143, 217, 353, 545, and 857 GHz) with this template. The extraction is performed within a 10°$\times$10° region centred on the X-ray cluster position, using a pixel size of 1.72$\times$1.72 arcmin$^2$. The signal is integrated up to $5R_{500}$ and then scaled back to $R_{500}$, according to the assumed pressure profile. The noise auto- and cross-spectra are computed directly from the data. This Python MMF implementation follows the same methodology and steps as the one implemented in \cite{planckcollaborationxxvii_planck_2016}. For additional details, we refer the reader to that paper and its references.

\subsection{$Y_{\text{SZ}}-M_\text{H}$ scaling relation}
\label{hydro_scaling}
To calibrate the $Y_{\text{SZ}}-M_\text{H}$ scaling relation, we follow the procedure described in Appendix A of \cite{planckcollaborationxx_planck_2014}. We invite the reader to refer to that work, as well as \cite{aymerich_cosmological_2024}, for the full details.

Given the fact that the ESZ sample is SNR limited, the clusters are expected to be biased high with respect to the mean near the detection threshold due to intrinsic scatter in the relation. To correctly retrieve the underlying scaling relation, this effect, known as Malmquist bias, is taken into account following the method introduced in \cite{vikhlinin_chandra_2009}.

Using the corrected $Y_{\text{SZ}}$ values, we calibrate the scaling relation between $Y_{\text{SZ}}$ and $M_{Y_{\text{X}}}$, the mass derived from the $Y_{\text{X}}-M_\text{H}$ relation, assuming the following form for the relation:\\
\begin{equation}
  \label{eq:Y_SZ-M_500_form}
  E^{-2/3}(z)\frac{D^2_A\,Y_{\text{SZ}}}{Y_{\text{piv}}}=10^{Y^*}\left(\frac{M_{Y_{\text{X}}}}{M_{\text{piv}}}\right)^{\alpha_\text{SZ}}.
\end{equation}
The data is fitted using a Markov Chain Monte Carlo (MCMC) approach with the \texttt{emcee} sampler \citep{foreman-mackey_emcee_2013}, by maximizing the following Gaussian log-likelihood with intrinsic scatter $\sigma_\text{int}$:
\begin{equation}
  \label{eq:likelihood}
  \begin{array}{l}
  \text{ln}\mathcal{L}=-0.5\sum_{i=1}^{N}\left[\frac{(y_i-(\alpha_\text{SZ} x_i+10^{Y^*}))^2}{2\sigma_i^2}-\text{ln}\left(\frac{1+10^{2 Y^*}}{2\pi\sigma_i^2}\right)\right],
  \end{array}
\end{equation}
where $x_i$ and $y_i$ are the logarithm of the mass and $Y_{\text{SZ}}$ values of the $i^{th}$ cluster, $\sigma_{x_{i}}$ and $\sigma_{y_{i}}$ are the uncertainties on the mass and $Y_{\text{SZ}}$ values, 
$\sigma_\text{int}$ is the intrinsic scatter of the relation, and $\sigma_i^2=\sigma_\text{int}^2+10^{2 Y^*} \sigma_{x_{i}}^2+\sigma_{y_{i}}^2$.

Finally, the uncertainties of the $Y_{\text{X}}-M_\text{H}$ are propagated to obtain the full $Y_{\text{SZ}}-M_\text{H}$ scaling relation:
\begin{equation}
  \label{eq:Y_SZ-M_500_Chandra}
  E^{-2/3}(z)\frac{D^2_A\,Y_{\text{SZ}}}{Y_{\text{piv}}}=10^{-0.29\pm0.01}\left(\frac{M_\text{H}}{M_{\text{piv}}}\right)^{1.70\pm0.1}.
\end{equation}

\subsection{Computation of the hydrostatic masses}
\label{hydro_mass_derivation}
We calculate the hydrostatic mass of all the clusters in the PSZ2 cosmological sample, using the scaling relation previously calibrated. To break the size-flux degeneracy, that is to make sure that the $Y_{\text{SZ}}$ signal is extracted in a $R_{500}$ radius corresponding to the final $M_\text{H}$ mass, we combine it with the $\theta_{500}-M$ relation from \cite{planckcollaborationxxvii_planck_2016}:
\begin{equation}
    \label{eq:theta_500_no_bias}
    \theta_{500}=\theta_{*} \left[\frac{h}{0.7}\right]^{-2/3} \left[\frac{M}{3.10^{14}M_{\odot}}\right]^{1/3} E^{-2/3}(z) \left[\frac{D_A(z)}{500 \, \text{Mpc}}\right]^{-1}.
  \end{equation}
We compute the masses via an MCMC approach to properly marginalize over both the posterior probability distribution in the $Y_{\text{SZ}}-\theta_{SZ}$ plane given in the \textit{Planck} catalogue and the uncertainties of the scaling relation parameters \citep[see Appendix A of][]{aymerich_cosmological_2024}.

\section{Weak-lensing calibration}
\label{bias}

\subsection{Shear profile likelihood}
We present here the general framework and derivation of the likelihood, before presenting how the different elements entering the likelihood are computed in Sect.~\ref{des_calibration}.

We extract radially binned tangential shear profiles $\hat{g}^i_\text{t}$ from the DES Y3 shape catalogue around every cluster in the DES calibration sample. From the \textit{Planck} data, we have redshifts $z^i$ and hydrostatic masses $M_\text{H}^i$ (derived in Sect.~\ref{hydro_masses}) for the full DES calibration sample. We denote $P(\hat{g}^i_\text{t}\vert M_{\text{WL}},z^i)$ the likelihood of observing a shear profile $\hat{g}^i_\text{t}$ given a weak-lensing mass $M_{\text{WL}}$ and a redshift $z^i$, and $P(M_{\text{WL}} \vert M,z^i)$ the likelihood of observing a weak-lensing mass $M_{\text{WL}}$ given an underlying halo mass $M$ and a redshift $z^i$. We note that $M_{\text{WL}}$ is a free parameter of the model that is marginalized over and not an actual mass derived for every cluster.

For a given cluster $i$, the likelihood of observing a tangential shear profile given a hydrostatic mass and a redshift is:\\
\begin{equation}
    \mathcal{L}_\text{WL}^i = P(\hat{g}^i_\text{t}\vert M_\text{H}^i,z^i) = \frac{P(\hat{g}^i_\text{t}, M_\text{H}^i \vert z^i) }{P(M_\text{H}^i \vert z^i)},    
\end{equation}
which can be written as:
\small
\begin{multline}
     \frac{P(\hat{g}^i_\text{t}, M_\text{H}^i \vert z^i) }{P(M_\text{H}^i \vert z^i)} = \\ 
     \frac{\int \text{d} M_{\text{WL}} \int \text{d} M P(\hat{g}^i_\text{t}\vert M_{\text{WL}},z^i) P(M_{\text{WL}} \vert M,z^i) P(M_\text{H}^i \vert M,z^i) P(\text{sel} \vert M_\text{H}^i,z^i) \frac{\text{d} N}{\text{d}M}}{\int \text{d} M P(M_\text{H}^i \vert M,z^i) P(\text{sel} \vert M_\text{H}^i,z^i) \frac{\text{d} N}{\text{d}M}},    
\end{multline}
\normalsize
where $P(\text{sel} \vert M_\text{H}^i,z^i)$ models the \textit{Planck} selection function and $P(M_\text{H}^i \vert M,z^i)$ the probability of obtaining a hydrostatic mass given a halo mass $M$ and a redshift $z^i$. The $P(\text{sel} \vert M_\text{H}^i,z^i)$ terms do not depend on the integrated variables, and therefore cancel out, and we introduce the usual prescription of a mass bias $(1-b)$ without scatter: 
\begin{equation}
    P(M_\text{H}^i \vert M,z^i) = \delta (M_\text{H}^i - (1-b) M),
\end{equation}
with $\delta(x)$ being the Dirac-delta function. This assumption, made following  \cite{planckcollaborationxx_planck_2014} and \cite{planckcollaborationxxiv_planck_2016}, can be justified by the very low scatter expected in the X-ray mass to halo mass relation compared to that of the WL mass to halo mass relation \citep{kravtsov_new_2006}.

The likelihood then simplifies as:
\begin{equation}
     \mathcal{L}_\text{WL}^i = \int \text{d} M_{\text{WL}} P(\hat{g}^i_\text{t}\vert M_{\text{WL}},z^i) P(M_{\text{WL}} \vert M=\frac{M_\text{H}^i}{1-b},z^i).
\end{equation}
The final log-likelihood is simply constructed from the sum of the individual ones:
\begin{equation}
     \text{ln}\mathcal{L}_\text{WL} = \sum_i \text{ln}\mathcal{L}_\text{WL}^i.
     \label{eq:mass_bias_likelihood}
\end{equation}
In our baseline analysis, we consider a fixed mass bias, but this modelling allows for the inclusion of a varying mass bias in a straightforward way, and we also explore the following mass bias parametrization introduced in \cite{salvati_mass_2019}:
\begin{equation}
  \label{eq:bias_salvati}
  1-b=\left( 1-b_0 \right) \left( \frac{M_\text{H}}{M_\text{piv}} \right) ^{\alpha} \left(\frac{1+z}{1+z_\text{piv}} \right)^{\beta},
\end{equation}
with $M_\text{piv} = 5 \cdot 10 ^{14} M_\odot$ and $z_\text{piv}=0.25$, corresponding approximately to the median values for the DES calibration sample.

\subsection{WL measurement and systematics assessment}
\label{des_calibration}

In the following, we describe the treatment and modelling of the WL data. We specify the cluster observables we measured in the WL data, the shear profile model used for the extraction of the mass information, and the simulation-based calibration of that mass extraction. These steps follow the analysis framework described in \citet{grandis_srg_2024} in the context of eROSITA, with adaptation to Planck.

\subsubsection{WL data products}
For each \textit{Planck} cosmology sample cluster $j$ in the DES Y3 footprint, we define 5 bins that are equally spaced in the log of the angular separation, corresponding to the physical scales $0.8 h^{-1}$~Mpc to $3.2/(1+z_{\text{cl},j})$, where $z_{\text{cl},j}$ is the cluster redshift, as well as one central bin from $0.5 h^{-1}$~Mpc to $0.8 h^{-1}$~Mpc. We query all DES WL sources falling in these bins, following the selection defined by \citet{myles_dark_2021}. To select sources in the background of each lens, we discard the first tomographic redshift bin and apply a weight to each source depending on the tomographic redshift bin $b$ it resides in, 
\begin{equation}\label{eq:tomo_weights}
    w^b = \begin{cases}
			\left\langle \Sigma_\text{crit,ls}^{-1} \right\rangle_{\text{s}\in b} &\text{for } z_\text{l} < z_{\text{med},b} \text{ and } b>1\\
            0. & \text{otherwise},
		 \end{cases}
\end{equation}
where the average is taken using the mean source redshift distribution of the tomographic bin, taken from \citet{myles_dark_2021}. The inverse critical lensing surface density expresses the geometrical configuration of the source $s$ and lens $l$ and is given by
 \begin{equation}\label{eq:sigmacritinv_ls}
     \Sigma_\text{crit,ls}^{-1} = \frac{4 \pi G}{c^2} \frac{D_\text{l}}{D_\text{s}} \max \left[0, D_\text{ls} \right],
 \end{equation}
where $G$ is the gravitational constant, $c$ the speed of light. Furthermore, we use the angular diameter distance between observer and source ($D_\text{s}$), and lens and source ($D_\text{ls}$).

Based on these sources, we compute the following WL observables:
\begin{itemize}
    \item the tangential reduced shear profile
    \begin{equation}\label{eq:cosmo_rdy_gt}
    \hat g_\mathrm{t} = \frac{\sum_{b=2,3,4}  w^b \sum_{i\in b}  w^\mathrm{s}_i e_{\mathrm t,\,b,i}  }{\sum_{b=2,3,4} w^b \sum_{i\in b} w^\mathrm{s}_i \mathcal{R}_{i} },
    \end{equation}
    where $w^\mathrm{s}$ is the source weight, $e_{\mathrm t}$ the tangential ellipticity w.r.t. the cluster position, and $\mathcal{R}$ the smoothed shear response \citep[see][ for more details]{grandis_srg_2024};
    \item the average distance of the sources from the cluster centres 
    \begin{equation}\label{eq:cosmo_rdy_theta}
  \theta =  \frac{\sum_{b=2,3,4} w^b  \sum_{i\in b}   w^\mathrm{s}_i \mathcal{R}_{i} \, \theta_i}{\sum_{b=2,3,4} w^b \sum_{i\in b} w^\mathrm{s}_i \mathcal{R}_{i} },
\end{equation} where $\theta_i$ is the angular separation between the source $i$ and the cluster;
\item the statistical uncertainty on the shear profile
\begin{equation}\label{eq:cosmo_rdy_dgt}
    \delta g_\mathrm{t} = \frac{\sigma_\text{eff}(z_\text{cl})}{\sqrt{N_\text{eff}}} \text{ with } N_\text{eff}=\frac{ \left( \sum_{b=2,3,4} \sum_{i\in b}   w^b  w^\mathrm{s}_i  \mathcal{R}_{i}  \right)^2}{ \sum_{b=2,3,4} \sum_{i\in b} \left(  w^b  w^\mathrm{s}_i \mathcal{R}_{i} \right)^2 },
\end{equation}
in the same bins, with the effective shape noise $\sigma_\text{eff}(z_\text{cl})$ estimated in \citet{grandis_srg_2024}, section~3.3.3;
\item  the radial source redshift distribution for fine source redshift bins with edges $(z_{\text{s}-}, z_{\text{s}+})$, reading
 \begin{equation}
     \hat P_\beta(z_{s}) = \frac{\sum_{b=2,3,4} w^b  \sum_{i\in b}   w^\mathrm{s}_i \mathcal{R}_{i} \, \text{I}_s(\hat z_{\beta,i})}{\sum_{b=2,3,4} w^b \sum_{i\in b} w^\mathrm{s}_i \mathcal{R}_{i} } \text{ for } \beta\in(\textsc{bpz},\textsc{dnf}),
 \end{equation}
 where $\text{I}_s( \hat z_{\beta,i})=1$ when $z_{\text{s}-}<\hat z_{\beta,i}<z_{\text{s}+}$. Practically speaking, this is a weighted histogram of the photo-$z$ estimates $\hat z_{\beta,i}$;
\end{itemize}

We also define a local background bin with physical separation between $7 h^{-1}$Mpc and $15h^{-1}$Mpc, in which we compute the field redshift distribution $P_\text{fld}(z_s)$ with self-organizing map redshift estimators \citep[following][Eq. 25]{grandis_srg_2024} for use in the mass extraction model (see Sect.~\ref{sec:extraction_model}) and with the \textsc{bpz} and \textsc{dnf} point estimates for use in the cluster member contamination calibration (see Appendix~\ref{cluster_member_contamination}).

All of the above observables are extracted in three different settings, based on the choice of the cluster centre: the SZ position, the cluster position found by an automated analysis of optical data \citep{hernandez-lang_pszmcmf_2023}, and the brightest central galaxies we determined.

\subsubsection{Shear extraction model}\label{sec:extraction_model}
We follow the WL mass calibration strategy proposed by \citet{grandis_calibration_2021} which calls for the definition of an easy-to-compute shear extraction model, which depends only on one free parameter, the so-called WL mass $M_\text{WL}$. As suggested by that work, we use a simplified mis-centred model based on a Navarro-Frenk-White profile \citep[NFW]{navarro_universal_1997}, corrected by the mean cluster member contamination. We keep the concentration of the profile fixed at $c_\text{500c}=3$. The radial density contrast of this model takes the form
\begin{equation}
    \Delta\Sigma(R | M) = \Delta\Sigma_\text{NFW}(R | M) - \left( \frac{R^{\text{extr}}_\text{mis}}{R}\right)^2 \Delta\Sigma_\text{NFW}(R^\text{extr}_\text{mis} | M),
\end{equation}
for $R>R^\text{extr}_\text{mis}$ and $\Delta\Sigma(R | M)=0$ else, where $R$ is the 2d-projected physical distance from the cluster centre. The mean mis-centring radius $R^\text{extr}_\text{mis}$ used for the mass extraction is defined based on the choice of observational centre,
\begin{equation}
    R^\text{extr}_\text{mis} = \begin{cases} 54.9^{\prime\prime} D_\text{l} [\text{Mpc}/^{\prime\prime} ] & \text{for SZ centred data,}\\
            78 \text{ kpc} & \text{for BCG centred data, and}, \\
            210 \text{ kpc} & \text{for MCMF centred data,}
		 \end{cases}
\end{equation}
which corresponds to the mean mis-centring in the three centring scenarios (see Table~\ref{table:centre}).

We then considered the average lensing efficiency 
\begin{equation}
    \Sigma_\text{crit,l}^{-1}=\left\langle \Sigma_\text{crit,ls}^{-1} \right\rangle_{P_\text{fld, SOMPz}(z_\text{s})}
\end{equation}
of the lens l,
obtained by averaging the lensing efficiency of the individual source, $\Sigma_\text{crit,ls}^{-1}$, Eq.~\ref{eq:sigmacritinv_ls}, with the background source redshift distribution $P_\text{fld, SOMPz}(z_\text{s})$ estimated with self-organizing maps. We employ this source redshift estimation method for the extraction model, as its source redshift distribution have undergone a rigorous calibration, as opposed to the empirical distributions based on \textsc{bpz} and \textsc{dnf}.

Using also the cluster member contamination fraction $f_\text{cl}(R | \lambda, z)$ (determined in Appendix~\ref{cluster_member_contamination}), evaluated at the mean cluster member contamination parameters, 
for the cluster richness $\lambda$ \citep[i.e. the number of member galaxies in the cluster, taken from][]{hernandez-lang_pszmcmf_2023} and the cluster redshift $z_\text{cl}$, we defined the reduced shear model:
\begin{equation}\label{eq:extraction_model}
    g_\text{t}^\text{mod}(R|M) = 
    \frac{\Sigma_\text{crit,l}^{-1}\Delta\Sigma(R | M)}{1-\Sigma_\text{crit,l}^{-1}\Sigma(R | M)} \left( 1-f_\text{cl}(R | \lambda, z_\text{cl})\right).
\end{equation}

The likelihood of the shear profile given the WL mass $M_\text{WL}$ then results from fitting the observed shear profile $\hat g_{\text{t}}$ with the extraction model evaluated at the cosmology-dependent position $R = D_\text{l}(z_\text{cl}) \theta$, reading 
\begin{equation}
    \ln P(\hat g_{\text{t}} | M_\text{WL}) =- \frac{1}{2}\left\{\hat g_{\text{t}}- g_\text{t}^\text{mod}(R|M_\text{WL})\right\}^\text{T}\mathsf{C}^{-1} \left\{\hat g_{\text{t}}- g_\text{t}^\text{mod}(R|M_\text{WL})\right\},
\end{equation}
where $\mathsf{C}$ is the covariance of the shear profile, composed of the squared shape noise $\delta g_\text{t}^2$ on the diagonal, added to a contribution from the uncorrelated large scale structure following \citet{hoekstra_how_2003}.

\subsubsection{WL mass to halo mass relation}
\label{wl_to_halo}
Given the intrinsic heterogeneity of clusters and the possible residual inaccuracies in the specified extraction model, the WL mass will be biased w.r.t. and scattered around the true halo mass, displaying a log-normal distribution \citep{grandis_calibration_2021}:
\begin{equation}
    P(\ln M_\text{WL} | M, z) = \mathcal{N}(\ln M_\text{WL}| \langle \ln M_\text{WL} | M, z\rangle, \sigma^2_\text{WL}(z)).
\end{equation}

The mean of the WL mass to halo mass relation is given by
\begin{equation}
\bigg< \ln \frac{M_\text{WL}}{M_\text{p}} \bigg| M, z_\text{cl} \bigg> = b(z_\text{cl}) + b_\text{M} \ln \left( \frac{M}{M_\text{p}} \right),
\label{eq:WLbias_given_M_z}
\end{equation}
with the pivot value for the mass  $M_\text{p} = 2 \cdot 10 ^{14} M_\odot$.

The redshift-dependent amplitude of the WL bias, its mass trend $b_\text{M}$, as well as the scatter $\sigma_\text{WL}(z)$ are calibrated from dedicated cluster WL simulations adapted from the prescription made in \citet{grandis_srg_2024}, section 4.4. to the context of \textit{Planck}-selected clusters. In Appendix~\ref{synth_shear_profiles}, we describe the simulation set-up in detail. Crucially, these simulations also provide systematic uncertainties of the WL bias, its mass trend, and the WL scatter, which we marginalize over during the parameter space exploration, by sampling the four parameters $A_\text{WL}$, $B_\text{WL}$, $C_\text{WL}$, and $D_\text{WL}$ defined in Appendix~\ref{synth_shear_profiles} within standard Gaussian priors.

We derive a set of WL bias and scatter numbers for each centring (BCG, SZ, MCMF) by adjusting both the mis-centring distribution in the simulation and adopting the corresponding extraction model. We vary the cluster member contamination determination between the results based on \textsc{bpz} and \textsc{dnf} (see Appendix~\ref{cluster_member_contamination}). Finally, we consider the scenario where we fit also for the inner bin between $0.5 h^{-1}$~Mpc to $0.8 h^{-1}$~Mpc, excluding it from the baseline analysis as it is the most sensitive to mis-centring problems \citep[see e.g.][]{sommer_weak_2024}. These permutations are summarized in Table~\ref{table:settings}, with the baseline analysis settings highlighted in bold (BCG centres, \textsc{bpz} source redshift measurement, excluding the inner bin and considering a fixed hydrostatic mass bias). In the following we refer to a given combination of analysis settings by its centre definition, inner bin radius (without units), source redshift measurement and adding the free bias evolution parameters when applicable, giving for example "BCG 800 $\textsc{bpz}$" for the baseline analysis or "BCG 800 $\textsc{bpz}$ $\alpha \beta$" when considering a mass bias evolving with both mass and redshift.
\begin{table}
  \caption{Summarry of analysis settings.}             
  \label{table:settings}      
  \centering                          
  \begin{tabular}{c c c c}        
  \hline\hline                 
  Centre & Source z & Inner bin & Prior on \\
   & measurement & radius & bias evolution \\
  \hline                        
  \textbf{BCG} & \textbf{\textsc{bpz}} & \textbf{800 kpc/h} & $\boldsymbol{\alpha=0}$\textbf{,} $\boldsymbol{\beta=0}$ \\
  MCMF & \textsc{dnf} & 500 kpc/h & $-5<\alpha<5$, $\beta=0$\\
  SZ &  & & $\alpha=0$, $-5<\beta<5$ \\
   & & & $-5<\alpha$, $\beta<5$ \\
  \hline                                   
  \end{tabular}
  \tablefoot{The top row, highlighted in bold, corresponds to the baseline analysis settings. Only one parameter is allowed to deviate from the baseline at once to limit the number of analyses (i.e. if studying evolving bias scenarios for example, the centres are BCG, the source redshift measurement method is \textsc{bpz} and the inner bin radius is 800 kpc/h).}
\end{table}
\section{Cluster abundance likelihood}
\label{cosmo}
To constrain the cosmological parameters $\Omega_{\text{m}}$ and $\sigma_8$ by fitting the observed cluster abundance, we base our analysis on the procedure from \cite{planckcollaborationxxiv_planck_2016}. This section is a brief summary of the number counts modelling likelihood, and the reader is referred to \cite{planckcollaborationxxiv_planck_2016} for more details.
\subsection{Number counts modelling}
\label{modelling}
In order to constrain the cosmological parameters \(\Omega_{\text{m}}\) and \(\sigma_8\) with cluster number counts, we need a prediction of theoretical cluster abundance as a function of cosmology. We choose the mass function from \cite{tinker_halo_2008} to predict the number of halos per unit mass and volume. We model the observed cluster number counts of the PSZ2 cosmological sample as a function of redshift $z$ (from follow-up observations) and SNR $q$, predicting the theoretical number counts with the following equation:
\begin{equation}
\frac{\text{d}N}{\text{d}z \, \text{d}q} = \int \text{d}\Omega_{\text{mask}} \int \text{d}M \, \frac{\text{d}N}{\text{d}z \, \text{d}M \, \text{d}\Omega} P[q \mid \bar{q}_{\text{m}}(M, z, l, b)],
\label{eq:dNdzdq}
\end{equation}
with \(\frac{\text{d}N}{\text{d}z \, \text{d}q}\) the theoretical prediction of cluster number counts, \(\frac{\text{d}N}{\text{d}z \, \text{d}M \, \text{d}\Omega}\) the mass function times the volume element, \(q\) the SNR, and \(\bar{q}_{\text{m}}(M, z, l, b)\) the mean SNR of a cluster with mass \(M\) at redshift \(z\). This term incorporates the scaling relation described in Sect.~\ref{hydro_masses} as well as the hydrostatic mass bias described in Sect.~\ref{bias} to link the mass and redshift of a cluster to its expected SNR:
\begin{equation}
    \bar{q}_{\text{m}} = \frac{\bar{Y}_{SZ}(M, z)}{\sigma_{\text{f}}[\bar{\theta}_{500}(M, z), l, b]},
\label{eq:qm}
\end{equation}
with \(\bar{Y}_{SZ}(M, z)\) the mean SZ signal of a cluster with mass \(M\) at redshift \(z\), coming from the scaling relation and mass bias, and \(\sigma_{\text{f}}\) the noise of the detection filter at the cluster's position \((l, b)\) and aperture \(\bar{\theta}_{500}\). For a given cosmology, the aperture is directly related to the mass and redshift through Eq.~\ref{eq:theta_500_no_bias}. 

The uncertainties and intrinsic scatter in the scaling relation, as well as the noise fluctuations and selection function of the survey are accounted for by the distribution \(P[q \mid \bar{q}_{\text{m}}]\). We use the analytical approximation from \cite{planckcollaborationxxiv_planck_2016} to model the survey's selection function, assuming pure Gaussian noise, leading to an error function form for the selection function.
\subsection{Number counts likelihood}
\label{likelihood}
Following \cite{planckcollaborationxxiv_planck_2016}, we construct the likelihood in the 2D space of redshift and SNR, dividing it into 10 redshift bins of width $\Delta z=0.1$ and 5 SNR 
bins of width $\Delta \text{log}q=0.25$. We model the observed cluster number counts $N(z_i,q_i)=N_{ij}$ in each bin as independent Poisson random variables with mean rates $\bar{N}_{ij}$, leading to the following log-likelihood:\\
\begin{equation}
  \label{eq:likelihood_cosmo}
  \text{ln}\mathcal{L}_\text{NC}=\sum_{i,j}\left[N_{ij} \text{ln}\bar{N}_{ij}-\bar{N}_{ij}-\text{ln}\left(N_{ij}!\right)\right].
\end{equation}
The theoretically predicted mean rates $\bar{N}_{ij}$ are obtained from Eq.~\ref{eq:dNdzdq} according to the cosmological parameters, scaling relation, and mass bias:\\
\begin{equation}
  \label{eq:mean_rates}
  \bar{N}_{ij}= \int_{z_i^-}^{z_i^+} \int_{q_j^-}^{q_j^+} \frac{\text{d}N}{\text{d}z \text{d}q} \text{d} z \text{d} q ,
\end{equation}
where $z_i^\pm$ and $q_j^\pm$ are the limits of the redshift and SNR bins.
\section{Constraining cosmological parameters}
Unlike the original \cite{planckcollaborationxxiv_planck_2016} study, that sampled the cosmology using priors from an external mass calibration on $Y^*$, $\alpha_\text{SZ}$, $\sigma_\text{int}$, and $(1-b)$, we do not use an informative prior on $(1-b)$ and instead combine the number counts and mass bias likelihood to constrain the hydrostatic mass bias simultaneously with the cosmology. The remaining parameters ($Y^*$, $\alpha_\text{SZ}$, and $\sigma_\text{int}$) are sampled within priors obtained from the X-ray mass calibration.
\subsection{Priors and sampling procedure}
 We use \texttt{Cobaya} Monte-Carlo-Markov-Chain sampler \citep{torrado_cobaya_2021} to sample the final likelihood combining the number counts likelihood (Eq.~\ref{eq:likelihood_cosmo}) and the mass bias likelihood (Eq.~\ref{eq:mass_bias_likelihood}). Following the original \cite{planckcollaborationxxiv_planck_2016} analysis, we also include the public \texttt{Cobaya} Baryon Acoustic Oscillation (BAO) likelihood with data from \cite{alam_clustering_2017} (BOSS DR12) in our sampling to constrain the value of $H_0$:
\begin{equation}
  \label{eq:total_likelihood}
  \text{ln}\mathcal{L}_\text{tot}=\text{ln}\mathcal{L}_\text{WL}+\text{ln}\mathcal{L}_\text{NC}+\text{ln}\mathcal{L}_\text{BAO},
\end{equation}
where $\text{ln}\mathcal{L}_\text{BAO}$ is the BAO log-likelihood. The impact of the BAO likelihood on our sampling is discussed in Appendix~\ref{BAO_impact}.

We consider the standard $\Lambda$CDM model, varying 5 parameters: the baryon density $\Omega_b h^2$, the cold dark matter density $\Omega_c h^2$, the angular size of the sound horizon at recombination $\theta_s$, the amplitude of the primordial power spectrum $A_s$, and the spectral index $n_s$. We use flat, uninformative priors for the mass bias $(1-b)$ and evolution parameters $\alpha$ and $\beta$ when considering an evolving bias. The $Y^*$, $\alpha_\text{SZ}$, and $\sigma_\text{int}$ parameters of the $Y_{\text{SZ}}-M_\text{H}$ scaling relation (see Eq.~\ref{eq:Y_SZ-M_500_form} and \ref{eq:Y_SZ-M_500_Chandra}) are sampled within the Gaussian priors obtained in Sect.~\ref{hydro_scaling}. We also introduce four parameters sampled within standard Gaussian priors to marginalize over the uncertainties on the WL mass to halo mass relation (see Sect.~\ref{wl_to_halo}).

Cluster number counts are a powerful cosmological probe to constrain the value of $\Omega_{\text{m}}$ and $\sigma_8$, but are not very sensitive to the rest of the cosmological parameters. Similarly to \cite{planckcollaborationxxiv_planck_2016}, we add a prior from Big Bang nucleosynthesis \citep[from][]{steigman_neutrinos_2008} for the value of $\Omega_b h^2= 0.0218\pm0.0012$ and a prior from CMB anisotropies \citep[from][]{planckcollaborationvi_planck_2020} for the value of $n_s=0.965 \pm 0.004$. We note that while this additional information is necessary to generate the halo mass function, its impact on the final $\Omega_{\text{m}}$ and $\sigma_8$ constraints is very limited. Table~\ref{table:priors} summarises the priors used to map the parameter space.
\begin{table}
  \caption{Priors on the parameters used in the MCMC sampling.}             
  \label{table:priors}      
  \centering                          
  \begin{tabular}{c c}        
  \hline\hline                 
  Parameter & Prior \\    
  \hline                        
  \multicolumn{2}{c}{Cosmology}\\
    $\theta_{\mathrm{s}}$ & $\mathcal{U}(0.01, 0.012)$\\
    $\log(10^{10} A_{\mathrm{s}})$ & $\mathcal{U}(2, 4)$\\
    $n_{\mathrm{s}}$ & $\mathcal{N}(0.965, 0.004)$\\
    $\Omega_{\mathrm{b}}h^2$ & $\mathcal{N}(0.0218, 0.0012)$\\
    $\Omega_{\mathrm{c}}h^2$ & $\mathcal{U}(0.01, 0.3)$\\
    $\sum m_\nu$ & $0.06$eV\\
  \hline                        
  \multicolumn{2}{c}{$Y_{\text{SZ}}-M_\text{H}$ scaling relation}\\
    $\alpha_{\text{SZ}}$ & $\mathcal{N}(1.70, 0.1)$\\
    $Y^*$ & $\mathcal{N}(-0.29, 0.01)$\\
    $\sigma_\text{int}$ & $\mathcal{N}(0.086, 0.01)$\\
  \hline                        
  \multicolumn{2}{c}{WL calibration}\\
    $(1-b)$ & $\mathcal{U}(0.3, 1.5)$\\
    $\alpha$ & 0 or $\mathcal{U}(-5, 5)$\\
    $\beta$ & 0 or $\mathcal{U}(-5, 5)$\\
    $A_\text{WL}$ & $\mathcal{N}(0, 1)$\\
    $B_\text{WL}$ & $\mathcal{N}(0, 1)$\\
    $C_\text{WL}$ & $\mathcal{N}(0, 1)$\\
    $D_\text{WL}$ & $\mathcal{N}(0, 1)$\\
     
  \hline                                   
  \end{tabular}
  \tablefoot{With $\mathcal{U}(\text{min}, \text{max})$ we indicate a uniform distribution between "min" and "max". With $\mathcal{N}(\mu, \sigma)$ we indicate a normal distribution centred on $\mu$ and with standard deviation $\sigma$. A single number indicates a fixed value. The prior used for $\alpha$ and $\beta$ varies between the different analysis settings (see Table~\ref{table:settings}).}
\end{table}
We report the constraints in terms of the cosmological parameters $\Omega_{\text{m}}$ and $\sigma_8$, derived from the underlying parameters $\Omega_b h^2$, $\Omega_c h^2$, $\theta_s$, $A_s$ and $n_s$.

\subsection{Cosmological constraints}
\label{constraints}
We derive cosmological constraints for different analysis settings, summarized in Table~\ref{table:settings}. We have different choices of halo centre definition, source redshift measurement, inclusion or exclusion of the inner bin, and prior on bias evolution. We do not explore all possible permutations, and choose to vary only one setting at a time, leaving the rest fixed to their baseline value given in the first row of Table~\ref{table:settings}.

\begin{figure}[]
    \centering
    \includegraphics[width=\columnwidth]{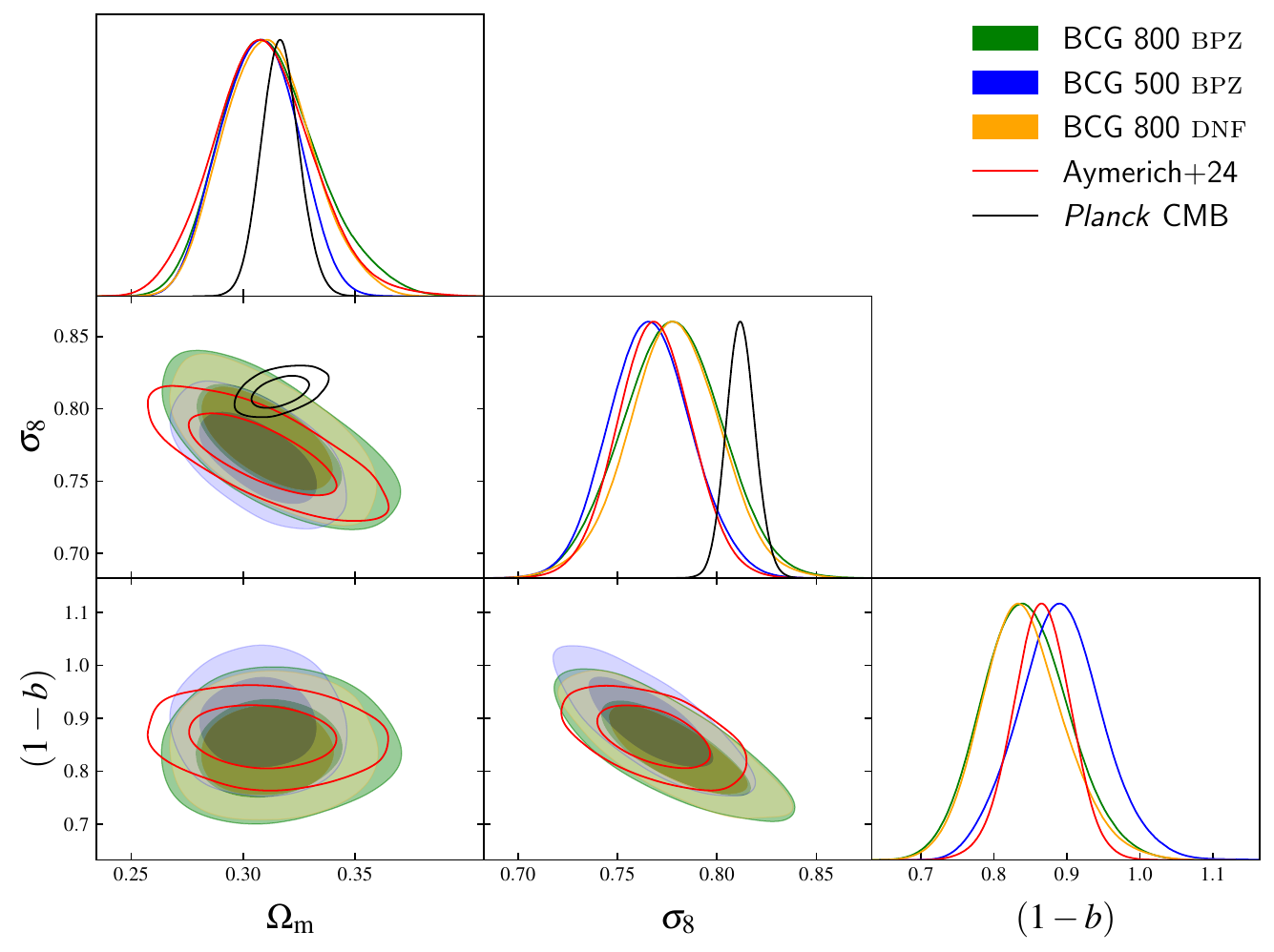}
    \caption{Final cosmological constraints obtained for the BCG 800 $\textsc{bpz}$, BCG 500 $\textsc{bpz}$ and BCG 800 $\textsc{dnf}$ analysis settings and comparison with constraints from SZ number counts obtained in \cite{aymerich_cosmological_2024} and constraints from CMB primary anisotropies from \cite{planckcollaborationvi_planck_2020}.}
    \label{fig:constraints_main}
\end{figure}

\begin{figure}[]
    \centering
    \includegraphics[width=\columnwidth]{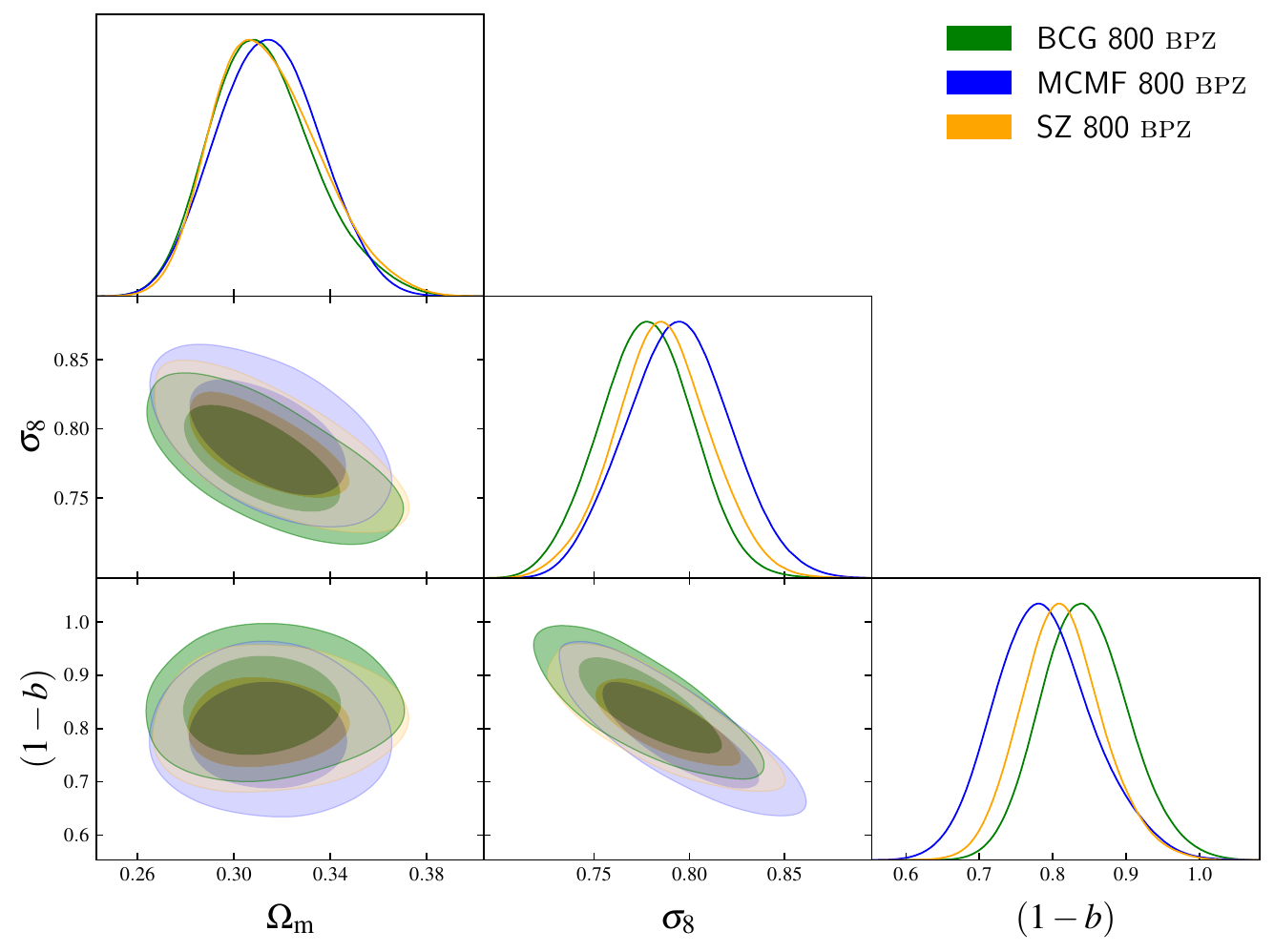}
    \caption{Final cosmological constraints obtained with the three different centre definitions used in this work (BCG, MCMF and SZ).}
    \label{fig:constraints_centre}
\end{figure}

\begin{figure}[]
    \centering
    \includegraphics[width=\columnwidth]{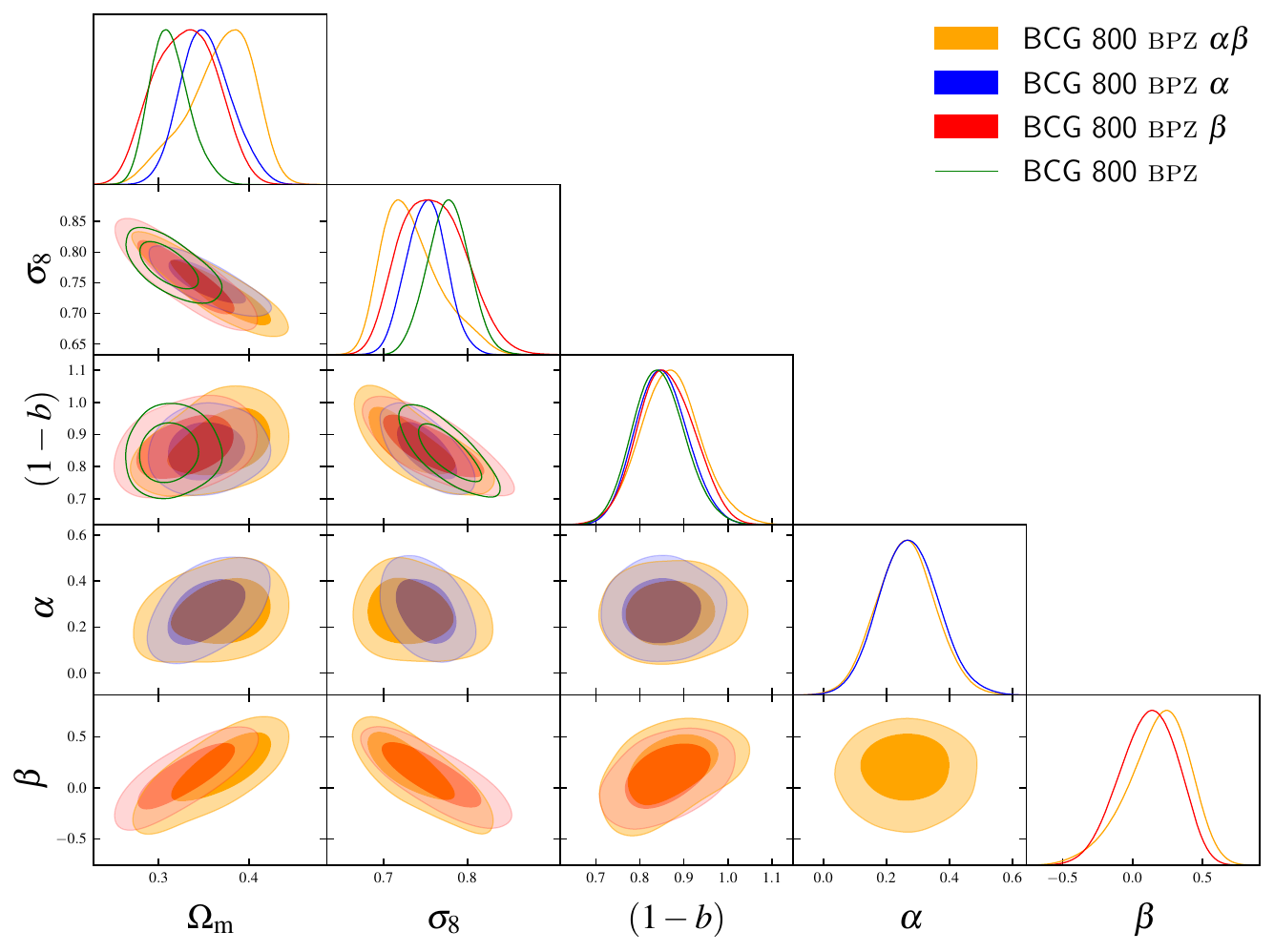}
    \caption{Final cosmological constraints obtained with the different bias evolution priors (fixed bias or evolution with mass and/or redshift). Fixing the bias evolution with mass (respectively redshift) is equivalent to setting $\alpha=0$ (respectively $\beta=0$).}
    \label{fig:constraints_evolv}
\end{figure}

Figure~\ref{fig:constraints_main} presents the constraints obtained when varying the source redshift measurement and including or excluding the inner bin. We also include, for comparison purposes, the constraints from \cite{aymerich_cosmological_2024}, obtained with the same cluster catalogue and X-ray data, but a different WL calibration sample, and the constraints from \cite{planckcollaborationvi_planck_2020}. Figure~\ref{fig:constraints_centre} presents the constraints obtained for the different centre definitions, and Fig.~\ref{fig:constraints_evolv} presents the constraints obtained with the various priors on bias evolution. Table~\ref{table:cosmo_2} reports the constraints on $\Omega_{\text{m}}$, $\sigma_8$, $S_8$, $(1-b)$, $\alpha$, and $\beta$ for all of the considered analysis settings, highlighting the baseline results in bold.

\begin{table*}
\caption{Constraints on $\Omega_{\text{m}}$, $\sigma_8$ and $S_8 \equiv \sigma_8 \sqrt{\Omega_\text{m} / 0.3}$, $(1-b)$, $\alpha$, and $\beta$ obtained for the different analysis settings combinations presented in Table~\ref{table:settings} and priors presented in Table~\ref{table:priors}.}             
\label{table:cosmo_2}      
\centering          
\begin{tabular}{c c c c c c c }     
\hline\hline       
Settings & $\Omega_{\text{m}}$ & $\sigma_8$ & $S_8$ & $(1-b)$ & $\alpha$ & $\beta$ \\ 
\hline                    
   \textbf{BCG 800 $\textsc{bpz}$} & $0.312^{+0.018}_{-0.024}$ & $0.777\pm 0.024$ & $0.791^{+0.023}_{-0.021}$ & $0.844^{+0.055}_{-0.062}$ & -- & -- \\
  MCMF 800 $\textsc{dnf}$ & $0.314\pm 0.021$ & $0.794\pm 0.027$ & $0.812^{+0.028}_{-0.026}$ & $0.789^{+0.059}_{-0.073}$ & -- & -- \\
  SZ 800 $\textsc{dnf}$ & $0.314^{+0.018}_{-0.025}$ & $0.786\pm 0.025$ & $0.803^{+0.022}_{-0.022}$ & $0.813^{+0.051}_{-0.059}$ & -- & -- \\
  BCG 500 $\textsc{bpz}$ & $0.307\pm 0.016$ & $0.767\pm 0.021$ & $0.774^{+0.020}_{-0.018}$ & $0.891\pm 0.058$ & -- & -- \\
  BCG 800 $\textsc{dnf}$ & $0.311^{+0.018}_{-0.021}$ & $0.778\pm 0.023$ & $0.792^{+0.020}_{-0.021}$ & $0.841^{+0.050}_{-0.061}$ & -- & -- \\
  BCG 800 $\textsc{bpz}$ $\alpha$ & $0.353^{+0.025}_{-0.031}$ & $0.751\pm 0.023$ & $0.814^{+0.019}_{-0.020}$ & $0.850^{+0.055}_{-0.062}$ & $0.272\pm 0.094$ & -- \\
  BCG 800 $\textsc{bpz}$ $\beta$ & $0.330\pm 0.034$ & $0.759^{+0.035}_{-0.043}$ & $0.792^{+0.021}_{-0.020}$ & $0.860\pm 0.064$ & -- & $0.12^{+0.22}_{-0.19}$ \\
  BCG 800 $\textsc{bpz}$ $\alpha \beta$ & $0.368^{+0.044}_{-0.027}$ & $0.734^{+0.024}_{-0.043}$ & $0.810 \pm 0.019$ & $0.870\pm 0.068$ & $0.262\pm 0.092$ & $0.18^{+0.26}_{-0.18}$ \\
\hline                  
\end{tabular}
\end{table*}

\section{Discussion}
\label{discussion}
\subsection{Coherence between analysis settings}
\begin{figure}[]
    \centering
    \includegraphics[width=\columnwidth]{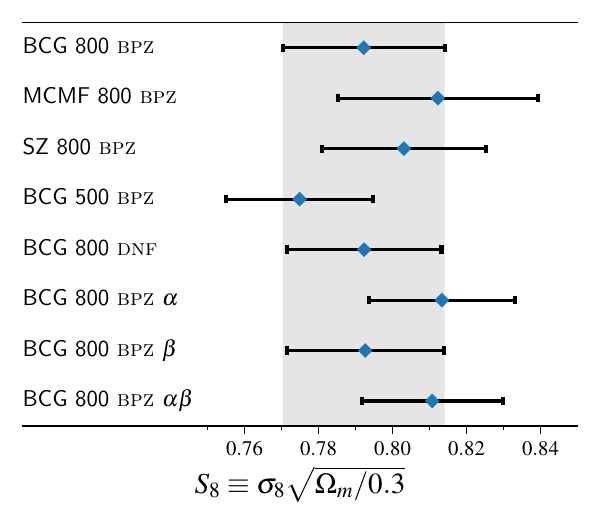}
    \caption{Comparison of the $S_8$ constraints (with 68\% confidence intervals) obtained with the different analysis settings combination.}
    \label{fig:S8_scenarios}
\end{figure}
Figure~\ref{fig:S8_scenarios} compares the $S_8 \equiv \sigma_8 \sqrt{\Omega_\text{m} / 0.3}$ obtained for all combinations of analysis settings to compare them and evaluate their coherence.
No combination of analysis settings leads to results in significant tension with the baseline analysis. With the exception of the scenarios with a bias evolving with mass, the results are even in very good agreement, deviating from the baseline by only a fraction of the 68\% confidence interval. This demonstrates that the analysis is robust to different modelling choices, and in particular, that the limited precision of the SZ centres and the presence of outliers in the optical centres can be systematically addressed by our shear extraction model and the weak-lensing mass–halo mass relation. 

We exclude the inner bin of the shear profiles from the baseline analysis given its higher sensitivity to mis-centring. Nevertheless, we note that the inclusion of the inner bin results in slightly tighter constraints and compatible preferred cosmological parameters, displaying once again the robustness of the WL modelling. 
 
The choice of using either \textsc{bpz} or \textsc{dnf} for the cluster member contamination correction results in essentially identical final constraints. This is likely due to a combination of the two measurements being reasonably coherent and to the exclusion of the inner radial bin, where cluster member contamination is most prevalent (see Fig.~\ref{fig:cluster_member_decomp}).

Letting the hydrostatic mass bias evolve with mass (i.e. not setting $\alpha=0$) leads to the largest deviation in constraints with respect to the baseline analysis (see Fig.~\ref{fig:constraints_evolv}). Nevertheless, we note that the constraints are shifted mostly along the $\Omega_{\text{m}} -\sigma_8$ degeneracy line, and that we obtain $S_8$ values compatible within 1\,$\sigma$ with the baseline regardless of the combination of analysis settings. The value and evolution of the hydrostatic mass bias are discussed in more details in Sect~\ref{mass_bias_discussion}.

\subsection{Goodness of fit}
\label{gof}
In this section, we evaluate the goodness of fit (GoF) of the various fits performed in the analysis. We do not present GoF results for the $Y_{\text{SZ}}-M_\text{H}$ scaling relation as they can be found in \cite{aymerich_cosmological_2024}, and focus on the GoF of the mass calibration with the WL shear profiles and of the binned number counts fit.

To evaluate the GoF of the shear profiles, we follow the approach of \cite{grandis_srg_2024} and stack tangential shear profiles for different bins in the sample. We choose to split the DES calibration sample into one low-mass bin ($M_\text{H}<4.5 \cdot 10 ^{14} M_\odot$) and three high-mass bins ($M_\text{H}>4.5 \cdot 10 ^{14} M_\odot$) with redshifts $z<0.23$, $0.23<z<0.34$ and $z>0.34$. The bin edges were chosen so that all four bins contain 23 or 24 clusters. We stack the tangential shear profiles of all the clusters $\varkappa$ in each of the bins, defining the stacked profiles $\hat g_{\text{t}}^\text{stack}$ and the error $ \delta\hat g_{\text{t}}^\text{stack}$ as:
\begin{equation}
  \label{eq:stacked_prof}
  \hat g_{\text{t}}^\text{stack}= \frac{\sum_\varkappa W_\varkappa \hat g_{\text{t},\varkappa}}{\sum_\varkappa W_\varkappa}
\end{equation}
and
\begin{equation}
  \label{eq:stacked_delta}
  \left(\delta\hat g_{\text{t}}^\text{stack}\right)^2= \frac{\sum_\varkappa W_\varkappa \delta \hat g^2_{\text{t}^,\varkappa}}{\sum_\varkappa W_\varkappa},
\end{equation}
where $W_\varkappa=\sum_{b=2,3,4} w^b \sum_{i\in b} w^\mathrm{s}_i \mathcal{R}_{i}$. The mean angular separation of the source galaxies is given by:
\begin{equation}
  \label{eq:stacked_angle}
  \theta^\text{stack}= \frac{\sum_\varkappa W_\varkappa \theta_\varkappa}{\sum_\varkappa W_\varkappa}.
\end{equation}

We compute the model prediction by drawing $N_\text{MC} = 1000$ points $p^\nu$ from the posterior sample. We compute the best-guess WL mass for each cluster $\varkappa$ and posterior point $\nu$ as the maximum of the PDF of WL masses given the cluster's hydrostatic mass and redshift:
\begin{equation}
  \label{eq:wl_mass_bestguess}
  M^\nu_{\text{WL},\varkappa}= \underset{M_{\text{WL}}}{\text{argmax }} P(M_{\text{WL}} \vert M_\text{H}^\varkappa,z^\varkappa,p^\nu).
\end{equation}
From each best-guess WL mass, we evaluate our extraction model $g_{\text{t},\varkappa}^{\text{mod},\nu}$ given in Eq.~\ref{eq:extraction_model}, stack them with the same weights as the data and compute the mean
\begin{equation}
  \label{eq:stacked_prof_t}
  g_{\text{t}}^\text{pred}= \frac{1}{N_\text{MC}} \sum_\nu \frac{\sum_\varkappa W_\varkappa g_{\text{t},\varkappa}^{\text{mod},\nu}}{\sum_\varkappa W_\varkappa},
\end{equation}
and variance
\begin{equation}
  \label{eq:stacked_delta_t}
  \left(\delta g_{\text{t}}^\text{pred}\right)^2 = \frac{1}{N_\text{MC}} \sum_\nu \frac{\sum_\varkappa W_\varkappa \left( g_{\text{t},\varkappa}^{\text{mod},\nu} - g_{\text{t}}^\text{pred} \right)^2}{\sum_\varkappa W_\varkappa},
\end{equation}
over the posterior points to propagate the uncertainty on the model parameters.
\begin{figure*}
    \centering
    \resizebox{0.99\hsize}{!}
    {\includegraphics{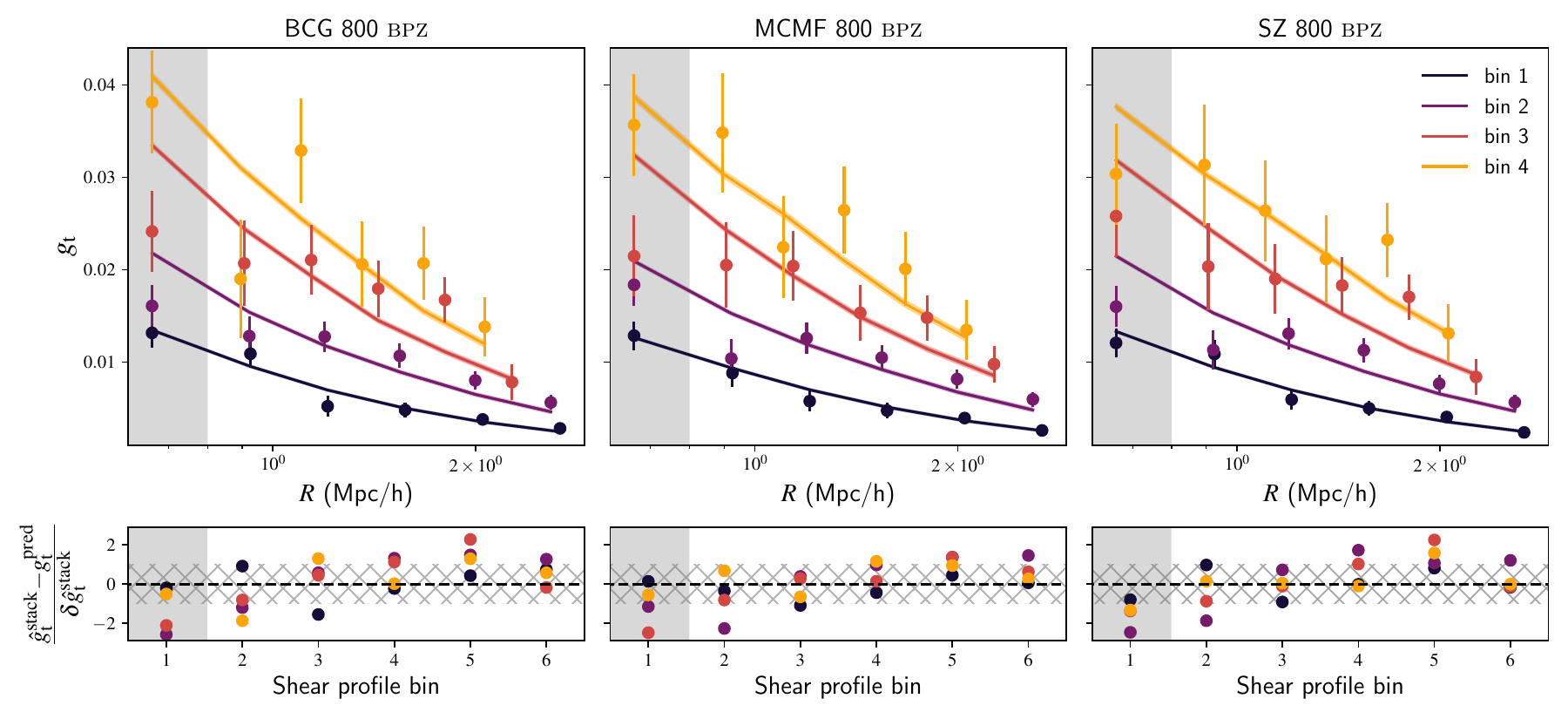}}
    \caption{\textbf{Top row}: Comparison of stacked tangential shear profile (data points) with the prediction from the best-fit model (lines), for the baseline analysis (left panel), MCMCF 800 $\textsc{bpz}$ analysis (central panel), and SZ 800 $\textsc{bpz}$ analysis (right panel). The shaded area corresponds to the inner bin, ignored in all analyses except BCG 500 $\textsc{bpz}$.\\
    \textbf{Bottom row}: Residuals rescaled by the uncertainty. The shaded area corresponds to the inner bin, and the hashed area to the 1$\sigma$ interval.}
    \label{fig:profiles}
\end{figure*}

Figure~\ref{fig:profiles} presents the stacked profiles and model predictions for the three centre definitions, with the rest of the settings set to their baseline value. The agreement is reasonably good (see the $\chi^2$ values reported in Table~\ref{table:GoF}), even if the observed profiles are on average a little flatter than the model predictions, especially for bins 2 and 3. While this difference in radial trend between the extraction model and the real data likely explains the slightly high $\chi^2$ values, it should not skew the recovered masses and hydrostatic mass bias. The WL mass is extracted with the correct weighting of the different radial scales, and its bias refers to the offset after averaging all radial scales. As such, our extraction model can systematically underestimate small scale shear while overestimating large scale shear due to an unrealistic choice of concentration and still remain unbiased. Assuming that the hydrodynamical simulations used for calibration have a realistic concentration, the WL mass bias $b_\text{WL}$ accounts for the difference and absorbs the potential bias coming from the use of an imperfect concentration in the extraction model. 

\begin{table}
  \caption{$\chi^2$ values of the shear profiles fitting and $C$ statistic of the number counts fitting for every combination of analysis settings.}             
  \label{table:GoF}      
  \centering                          
  \begin{tabular}{c c c}        
  \hline\hline                 
  Analysis settings & $\chi^2_\text{shear}/N_\text{DoF}$ & $C_\text{NC}$\\    
  \hline                        
  BCG 800 $\textsc{bpz}$ & $26.0^{+2.1}_{-2.7}/17$ & $46.4$\\
  MCMF 800 $\textsc{bpz}$& $18.3^{+1.5}_{-2.1}/17$ & $46.4$\\
  SZ 800 $\textsc{bpz}$ & $21.6^{+1.8}_{-2.3}/17$ & $46.4$\\  
  BCG 500 $\textsc{bpz}$ & $37.4^{+1.4}_{-1.9}/21$ & $46.6$\\
  BCG 800 $\textsc{dnf}$ & $26.0^{+2.1}_{-2.6}/17$ & $46.3$\\
  BCG 800 $\textsc{bpz}$ $\alpha$ & $25.8^{+2.2}_{-2.7}/16$ & $47.9$\\
  BCG 800 $\textsc{bpz}$ $\beta$ & $25.9^{+2.2}_{-2.7}/16$ & $46.8$\\
  BCG 800 $\textsc{bpz}$ $\alpha \beta$ & $25.7^{+2.1}_{-2.6}/15$ & $46.7$\\
  \hline                                   
  \end{tabular}
  \tablefoot{The expected value of the C-statistic, given the considered models, is $C_\mathrm{NC}=34.5\pm 7.9$.}
\end{table}

\begin{figure}[]
    \centering
    \includegraphics[width=\columnwidth]{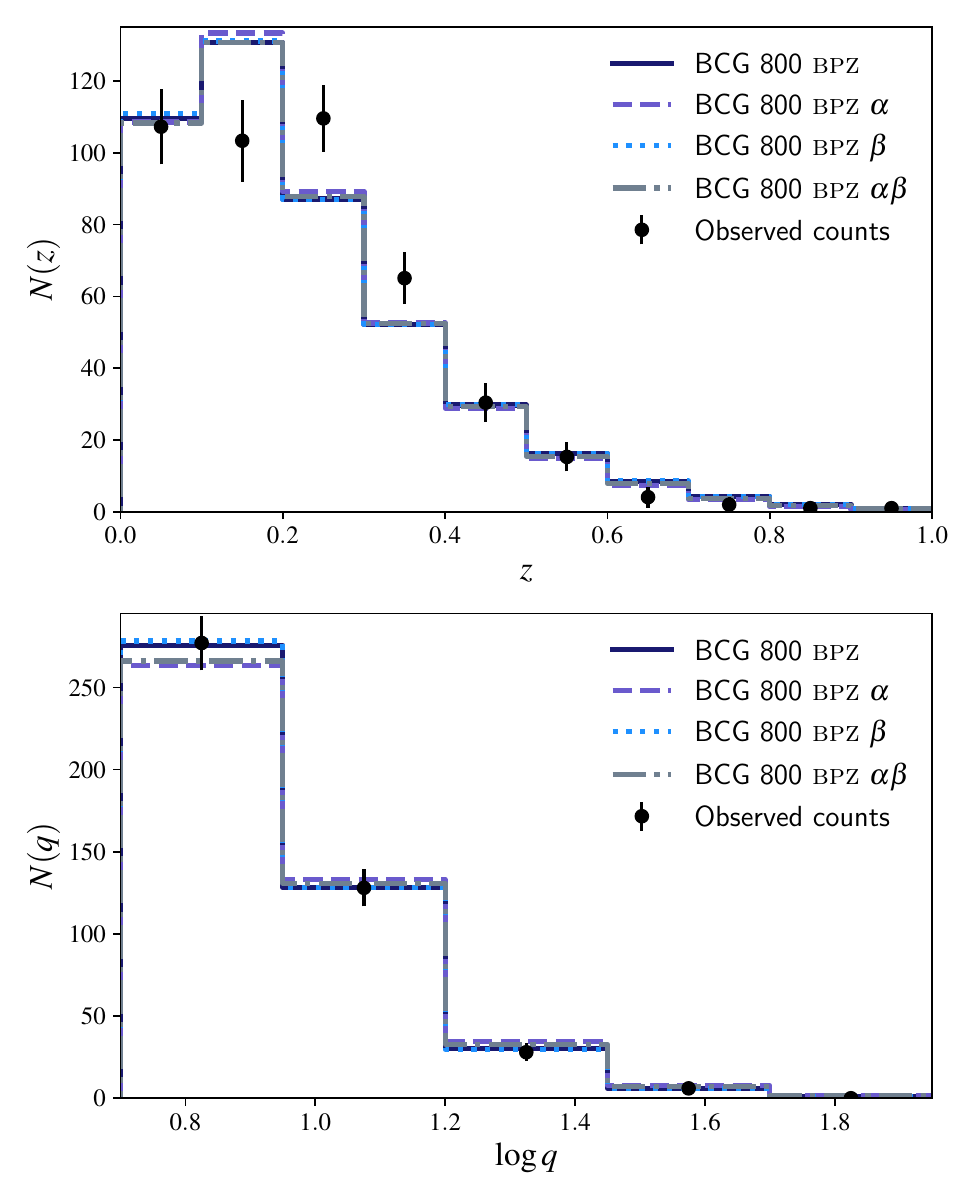}
    \caption{Comparison of the observed number counts of the PSZ2 sample (data points) with the prediction from the best-fit model for the baseline analysis and the different evolving bias scenarios (histograms). The error bars correspond to the $\sqrt{N}$ standard deviation expected from a Poisson distribution with parameter $N$.}
    \label{fig:GoF_NC}
\end{figure}

We also investigate the GoF of the number counts fitting, by comparing the observed number counts with the predicted cluster abundance from the maximum likelihood model. To evaluate the GoF, we choose to use the $C$ statistic \citep[see e.g.][]{bonamente_distribution_2020}, as it is the most relevant statistic in the case of data fitted with a Poisson likelihood:
\begin{equation}
  \label{eq:c_stat}
  C= 2 \sum_{i,j} \bar{N}_{ij} - N_{ij} + N_{ij}\ln(N_{ij} / \bar{N}_{ij}),
\end{equation}
where $\bar{N}_{ij}$ and $N_{ij}$ are the predicted and observed binned number counts, as defined in Eq.~\ref{eq:mean_rates}.
The $C$ statistic values found for all the analysis settings combinations are reported in Table~\ref{table:GoF}. We also visually present the fit in Fig.~\ref{fig:GoF_NC} for the baseline analysis, as well as the analyses with an evolving bias, as it is the only setting that has a noticeable impact on the number counts fit.

The interpretation of the C-statistic is not as straightforward as that of the $\chi^2$ statistic because its expected distribution is model-dependent. We evaluate it via a Monte-Carlo approach, by generating 10\,000 Poisson distributions from the best-fit model and computing the C-statistic. We found that for all of the analysis settings, the expected value of the C-statistic is $C_\mathrm{NC}=34.5\pm 7.9$. We thus find that the fit of the number counts is reasonably good, but not ideal. This is a well-known problem of the PSZ2 sample, whose redshift distribution does not match halo mass functions very well \citep[see Fig.~6 of][]{planckcollaborationxxiv_planck_2016}. Since this work only impacts the mass calibration of the PSZ2 sample, its redshift distribution is unchanged and this problem remains.

We note that neither the shear profiles nor number counts GoF, as measured by the $\chi^2$ and $C$ statistics, are improved by allowing the bias to evolve with mass (i.e. freeing $\alpha$), which is the only setting change that leads to a significant shift in cosmological constraints. To further investigate why this bias evolution leads to a significant departure from the self-similar value, we present the maximal value of the likelihood $\text{ln}\hat{\mathcal{L}}_\text{tot}$ for the baseline setting and the two analyses with a bias evolving with mass in Table~\ref{table:likelihood}. We also compute the difference in Akaike Information Criteria (AIC) with respect to the baseline analysis $\Delta_\text{AIC}$, with the AIC being defined as:
\begin{equation}
  \label{eq:aic}
  \text{AIC} = -2 \, \text{ln}\mathcal{L}_\text{tot} + 2\,\text{dim}(\mathcal{M}),
\end{equation}
where $\text{dim}(\mathcal{M})$ is the number of model parameters in the model $\mathcal{M}$. The AIC allows for a quantitative measurement of whether the improvement in the fit justifies the inclusion of additional model parameters, following the principle known as Ockham's razor (limiting the number of parameters to the necessary minimum).

\begin{table}
  \caption{Maximal $\text{ln}\hat{\mathcal{L}}_\text{tot}$, separated into $\text{ln}\hat{\mathcal{L}}_\text{NC}$, $\text{ln}\hat{\mathcal{L}}_\text{WL}$, and $\text{ln}\hat{\mathcal{L}}_\text{BAO}$, and $\Delta_\text{AIC}$ for the baseline analysis and the analysis with $\alpha$ left free to vary.}             
  \label{table:likelihood}      
  \centering                          
  \begin{tabular}{c c c c}        
  \hline\hline                 
  & BCG 800 $\textsc{bpz}$ & BCG 800 $\textsc{bpz}$ $\alpha$ & BCG 800 $\textsc{bpz}$ $\alpha \beta$ \\    
  \hline                        
  $\text{ln}\hat{\mathcal{L}}_\text{NC}$ & $-68.22$ & $-68.44$ & $-68.49$\\
  $\text{ln}\hat{\mathcal{L}}_\text{WL}$ & $-272.38$ & $-269.12$ & $-268.50$\\
  $\text{ln}\hat{\mathcal{L}}_\text{BAO}$ & $32.40$ & $34.04$ & $34.09$\\
  $\text{ln}\hat{\mathcal{L}}_\text{tot}$ & $-308.20$ & $-303.52$ & $-302.90$\\
  $\Delta_\text{AIC}$ & $0$ & $7.36$ & $6.60$\\  
  \hline                                   
  \end{tabular}
\end{table}

We find that the inclusion of $\alpha$ as a free parameter leads to a significant enough improvement in fit of both WL and BAO data, while leaving the number counts fit almost unchanged, to be favoured by the AIC. On the other hand, also freeing $\beta$ does not significantly improve the fit to justify its inclusion with the AIC.

\subsection{Comparison with other cosmological analyses}
\label{other_probes}
\begin{figure}[]
    \centering
    \includegraphics[width=\columnwidth]{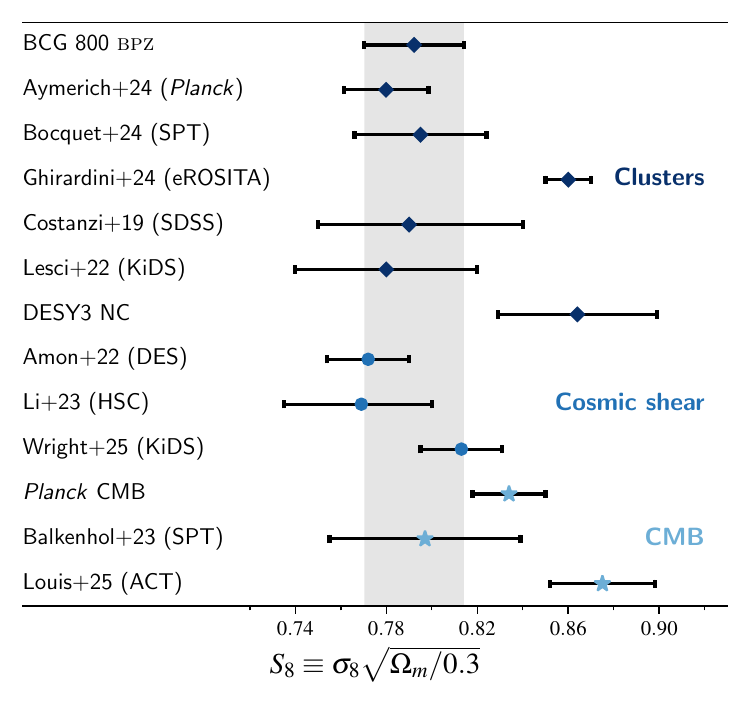}
    \caption{Comparison of the $S_8$ constraints (with 68\% confidence intervals) obtained in the baseline analysis of this work with other recent analyses.}
    \label{fig:S8_probes}
\end{figure}
Figure~\ref{fig:S8_probes} compares the $S_8 \equiv \sigma_8 \sqrt{\Omega_\text{m} / 0.3}$ constraints obtained in this work with those from some recent cosmological analyses using both cluster number counts as well as other different cosmological probes. For the comparison to be relevant, we only selected analyses with the same $S_8$ definition, which might not be optimal for every study, in the sense that another definition might better match their $\sigma_8 -\Omega_\text{m}$ degeneracy.

We compare our baseline analysis with results from \cite{aymerich_cosmological_2024}, \cite{bocquet_spt_2024a} (SPT SZ cluster number counts), \cite{ghirardini_srg_2024} (eROSITA cluster number counts), \cite{costanzi_methods_2019} (SDSS cluster number counts), \citep{lesci_amico_2022} (KiDS DR3 cluster number counts), \cite{descollaboration_dark_2025} (DES Y3 cluster number counts), \cite{amon_dark_2022} (DES Y3 cosmic shear), \cite{wright_kidslegacy_2025} (KiDS Legacy cosmic shear), \cite{li_hyper_2023} (HSC Y3 cosmic shear), \cite{planckcollaborationvi_planck_2020} (noted \textit{Planck} CMB), \cite{louis_atacama_2025} (ACT DR6 CMB),
\cite{balkenhol_measurement_2023} (SPT-3G CMB). For clarity of the plot, we chose one HSC result amongst four analyses, 
but \cite{dalal_hyper_2023}, \cite{miyatake_hyper_2023}, and \cite{sugiyama_hyper_2023} find very similar results to the one chosen for the plot.

We find that our work results in a slightly low $S_8$ value compared to most CMB primary anisotropies analyses (the tension with \textit{Planck} results is very moderate at 1.6$\sigma$)\footnote{all tensions mentioned in this work are computed as $\Delta \mu / \sqrt{\sigma_1^2 + \sigma_2^2}$, with $\Delta \mu$ the difference in central values and $\sigma_i$ the respective uncertainties.}, and in general agreement with most late-time probe results (clusters and cosmic shear). In particular, we find results in great agreement with the previous study of the \textit{Planck} sample done in \cite{aymerich_cosmological_2024}, despite changing the WL data used for the mass calibration. 

We find a lower $S_8$ value than the latest study of optically detected DES clusters presented in \citet{descollaboration_dark_2025}, although the difference is not statistically significant given the uncertainties. It is worth noting that while this study includes cluster number counts, it also includes additional information in the form of two-point correlation functions derived from galaxy density, WL shear, and cluster density. The constraints presented in Fig.~\ref{fig:S8_probes} were derived using cluster abundances and four two-point correlation functions, and the addition of cosmic shear and galaxy-galaxy lensing to the analysis shifts the constraints to $S_8=0.811^{+0.022}_{-0.020}$, in good agreement with our results.

Our $S_8$ value is in perfect agreement, with similar constraining power, with the results from \cite{bocquet_spt_2024a}, a study of SPT SZ-selected clusters with mass calibration from DES data. The fact that we obtain slightly tighter constraints than \citet{bocquet_spt_2024a}, despite having roughly half the number of clusters, can be explained by five reasons. First, while the sample is smaller, the constraining power of cluster studies is currently limited by the mass calibration and not the statistics of the cluster sample, as shown in Sect.~\ref{forecast}. Despite having fewer clusters in the WL calibration sub-sample, the \textit{Planck} clusters are on average heavier and at lower redshift than the SPT clusters, leading to a higher average WL SNR per cluster. Secondly, our study incorporates X-ray data, which acts as a low-scatter proxy that constrains the scatter and slope of the $Y_\text{SZ}$ to mass relation. Third, the SPT analysis marginalizes over two parameters that correspond to a mass and redshift evolution of their mass-observable relation, whereas we use a fixed mass bias in the case of our baseline analysis, and only marginalize over free mass dependence at the level of the $Y_{\text{SZ}}-M_\text{H}$ scaling relation. Fourth, we also simultaneously sample the BAO likelihood, which constrains the value of $H_0$, helping with the mass calibration, and slightly constrains the lower bound of $\Omega_\mathrm{m}$, unlike the SPT analysis which uses a Gaussian prior on $H_0$ that is wider than the posterior we obtain (see Appendix~\ref{BAO_impact}). Finally, reporting the constraining power in terms of $S_8$ is unfavourable to the SPT analysis as the degeneracy they obtain aligns better with $\sigma_8 (\Omega_\mathrm{m}/0.3)^{0.25}$ rather than $S_8 \equiv \sigma_8 \sqrt{\Omega_\text{m} / 0.3}$.

\begin{figure}[]
    \centering
    \includegraphics[width=\columnwidth]{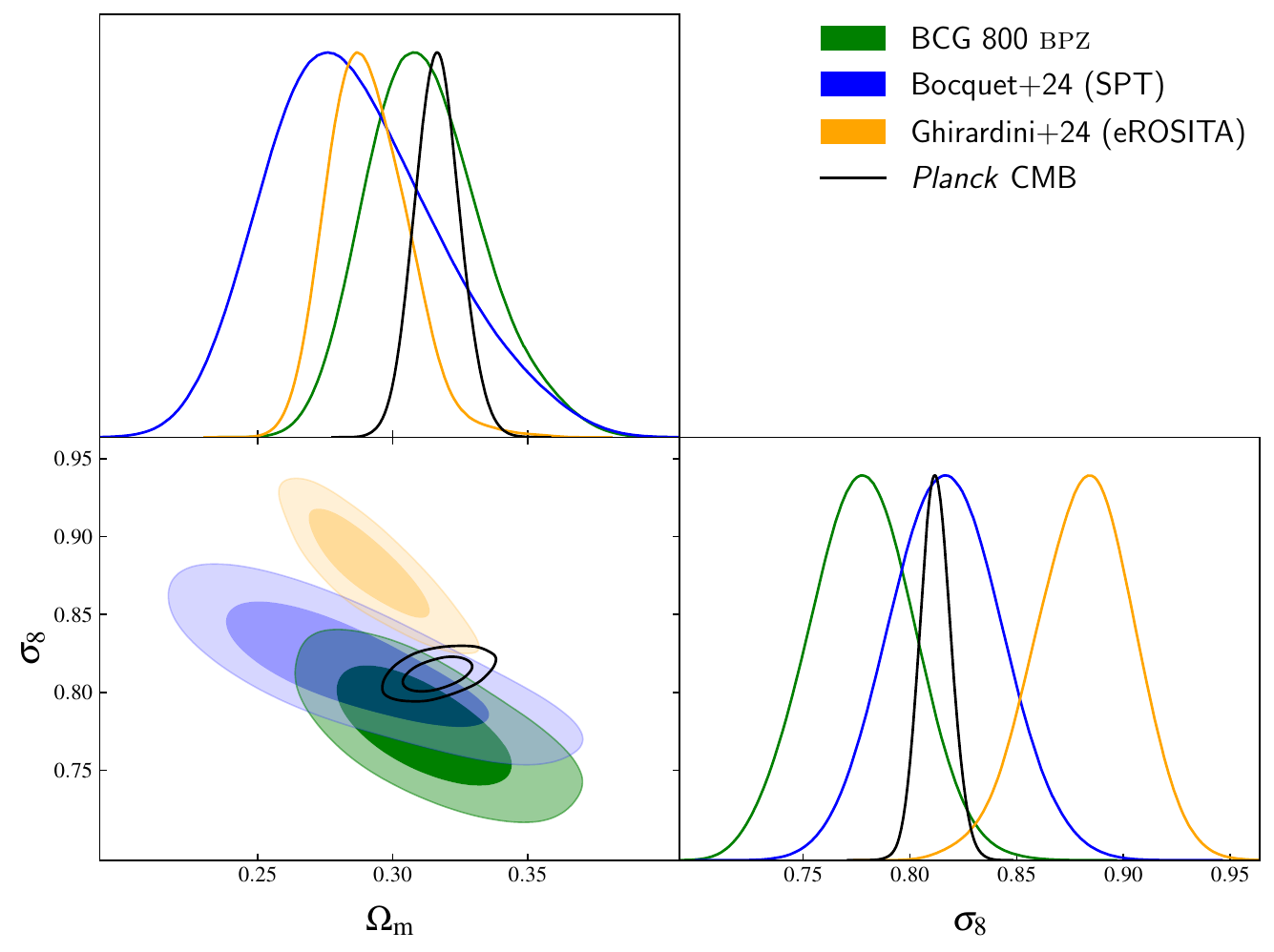}
    \caption{Comparison of the constraints obtained in the baseline analysis of this work with the results from \cite{bocquet_spt_2024a}, \cite{ghirardini_srg_2024}, and \cite{planckcollaborationvi_planck_2020}. We note that our work includes BAO data, improving the mass calibration by constraining $H_0$ and helping to constrain $\Omega_\mathrm{m}$ slightly (see Appendix
    ~\ref{BAO_impact}), while the SPT analysis adopts a wide prior on $H_0$ and the eROSITA analysis adopts a tight prior on $H_0$.}
    \label{fig:constraints_comp}
\end{figure}

We find a significant tension of 2.9$\sigma$ between our results and those of \cite{ghirardini_srg_2024} (eRASS1 cluster cosmology, with DES WL mass calibration), despite sharing the same mass calibration data and procedure. This tension is reduced slightly to 2.3$\sigma$ when considering a mass bias evolving with mass and redshift (BCG 800 $\textsc{bpz}$ $\alpha \beta$ analysis). To better understand the tensions between this work, \cite{bocquet_spt_2024a}, and \cite{ghirardini_srg_2024}, we compare the constraints in the $\Omega_{\text{m}} -\sigma_8$ plane in Fig.~\ref{fig:constraints_comp}, as the $S_8$ value only gives a limited picture of possible tensions. This comparison confirms the good agreement between this work and \cite{bocquet_spt_2024a}, even if the latter prefers a slightly higher $\sigma_8$ value and slightly lower $\Omega_\text{m}$. The tension between this work and \cite{ghirardini_srg_2024} is also highlighted, with no overlap of the 2$\sigma$ contours. 

\begin{figure}[]
    \centering
    \includegraphics[width=\columnwidth]{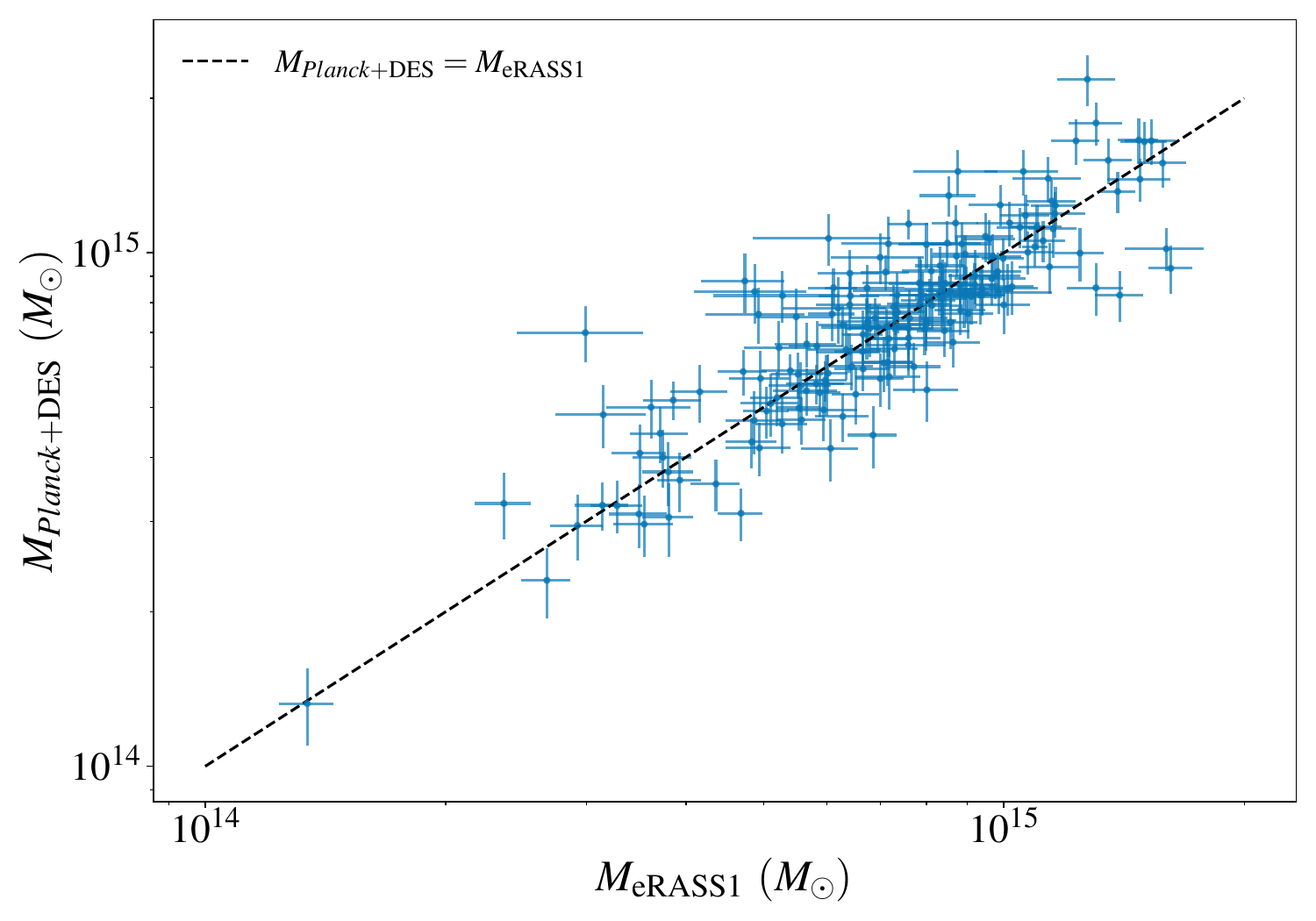}
    \caption{Comparison of eRASS1 and \textit{Planck}+DES mass for all PSZ2 clusters with a match in the eRASS1 catalogue (see text for the definition of the masses). The dashed line corresponds to the unity line and is a simple visual guide, not a fit.}
    \label{fig:erosita_mass}
\end{figure}

To confirm that the mass calibration is coherent between both studies and cannot explain the tension, we compare the \textit{Planck}+DES mass with the eRASS1 mass for every PSZ2 cluster with a match in the eRASS1 catalogue \citep{bulbul_srg_2024} in Fig.~\ref{fig:erosita_mass}. For the eRASS1 mass, we directly take the mass and uncertainty from the eRASS1 catalogue, computed using the best-fit cosmology and scaling relations from \cite{ghirardini_srg_2024}. For the \textit{Planck}+DES mass, we take the hydrostatic masses derived in Sect.~\ref{hydro_mass_derivation} and divide them by the best-fit bias of our baseline analysis:
\begin{equation}
\label{eq:planck_des_mass}
M_{\textit{Planck}\text{+DES}}=\frac{M_\text{H}}{(1-b)}.
\end{equation}
The vertical error bars in Fig.~\ref{fig:erosita_mass} account for the uncertainty on both the hydrostatic masses and bias.

We find an excellent agreement between the masses derived from eROSITA and \textit{Planck} data. This might be expected, as both experiments use the same WL data, and treat it in very similar ways. Yet, both experiments calibrate very different mass proxies: X-ray photon count rate for eRASS1 and the SZ signal informed by X-ray information in the case of Planck. The respective mass calibration procedures make very different assumptions on the scaling relations, the completeness and selection function modelling, and the treatment of sample purity. Agreement between the mass estimates is thus far from a given, and validates that these analysis choices do not bias the mass calibration. Yet, they also lead to different cosmological results from the respective number counts. This means that the difference in final cosmological constraints is likely coming from the difference in catalogue and population model (i.e. selection function, completeness evaluation, and likelihood set-up). Identifying the actual reasons for the tension between our results and the eRASS1 cosmology is beyond the scope of this work but is likely to reveal crucial aspects of cluster population modelling in a cosmological context.

\subsection{Derived value of the hydrostatic mass bias}
\label{mass_bias_discussion}
In our baseline analysis, we obtain $(1-b)=0.844^{+0.055}_{-0.062}$ and find no significant departures from this value regardless of the analysis settings. While it is in general agreement with most studies of the hydrostatic mass bias, comparisons of different analyses are not straightforward and should be done case by case. Indeed, the mass bias measured in this study and most other observational derivations of $(1-b)$ is not a pure hydrostatic mass bias as can be measured in simulations, but rather an X-ray mass bias. While the deviations from hydrostatic equilibrium do impact the value of $(1-b)$, other observational factors also come into play. The main residual effect is the calibration of the X-ray instrument, which is a well-known issue in X-ray astronomy. While this has no impact on the final cosmological constraints due to the WL mass calibration, it significantly changes the value of the mass bias \citep{aymerich_cosmological_2024}. In this work, we use X-ray data from the \textit{Chandra} telescope, which does not measure the same temperature and therefore hydrostatic mass as, for example, XMM-\textit{Newton}. This difference in calibration is expected to yield a final $(1-b)$ value $\sim16\%$ higher when using \textit{Chandra} data \citep{schellenberger_xmmnewton_2015, potter_hydrostatic_2023, aymerich_cosmological_2024}.

This calibration problem does not appear in results derived from simulations and thus has to be treated as an unknown systematic shift when comparing our results with simulations. Most results from simulations find a mass bias in the $[0.8,0.9]$ range, in excellent agreement with our value \citep[see][for a review of many derivation of $(1-b)$ from simulations]{gianfagna_exploring_2021}. 

On the observational side, we can compare our results with two recent studies that derived a mass bias on a subsample of \textit{Planck} clusters: \citet{wicker_constraining_2023} and \citet{sereno_chexmate_2025}. The first study, based on gas mass fraction, is not expected to be very sensitive to the calibration problem, as the temperature did not enter the analysis. The value of $(1-b)=0.842\pm0.040$ derived in \citet{wicker_constraining_2023} is in excellent agreement with our study. The second study derived a bias by comparing dynamical masses (that they found to be unbiased compared to WL masses) to \textit{Planck} masses, that are obtained via a scaling relation calibrated with XMM-\textit{Newton} hydrostatic masses. \citet{sereno_chexmate_2025} derived a lower bias value than our result, finding $(1-b)=0.62\pm0.04$. This time, the calibration inconsistency between XMM-\textit{Newton} and \textit{Chandra} needs to be taken into account, and their result has to be rescaled by $15\%$ to a \textit{Chandra}-like bias to allow for a proper comparison, yielding $(1-b)_\mathrm{scaled}=0.72\pm0.05$. This bias is still significantly lower than the result from this work, even though the tension is reduced when accounting for instrumental effects. Interestingly, they found conclusive evidence for a $(1-b)$ value increasing with mass (i.e. heavier clusters showing less deviation from hydrostatic equilibrium), similarly to what is obtained in this study when freeing the value of $\alpha$.

Finding deviations from the self-similar value when opening up the bias evolution with mass with the \textit{Planck} cluster sample is not new, and the trend is known to vary when considering sub-samples \citep[see e.g.][]{salvati_mass_2019, wicker_constraining_2023}. The large correlation between redshift and cluster mass in the PSZ2 sample (see Fig.~\ref{fig:m_z}) is likely contributing to this problem, making the separation of mass and redshift trends challenging. 

\subsection{Next-generation forecast}
\label{forecast}
The results of this work demonstrate the clear potential of cluster mass calibration in wide photometric surveys designed for cosmic shear experiments, like DES, as opposed to deep, targeted observations of individual clusters. The current results based on DES-Y3 lead to a mass calibration comparable to studies based on targeted WL \citep[e.g.][]{aymerich_cosmological_2024}. This is, however, going to change with the advent of Stage IV WL surveys such as $\textit{Euclid}$, LSST-\textit{Rubin}, and \textit{Roman}.

Following the forecast framework for LSST-\textit{Rubin} WL presented in \citet{grandis_impact_2019}, section 4.2, we create synthetic, LSST-\textit{Rubin}-like shear profiles for the 266 \textit{Planck} clusters with $\text{DEC}< 15$ deg. In short, we assume a source number density of  $40$ arcmin$^{-2}$ instead of $5$-$6$ arcmin$^{-2}$ for DES-Y3\footnote{Our estimate for the WL source number density is taken from the LSST science book, section 3.7.2, \url{https://www.lsst.org/sites/default/files/docs/sciencebook/SB_3.pdf}. This estimate is very optimistic, as it does not consider the impact of blending and masking \citep{chang13}.}. The median source redshift distribution is assumed to be $z_\text{s}=0.8$. We assume a WL bias $b_\text{WL}=1\pm 0.01$, and a hydrostatic mass bias $b=0.8$, as well as a WL scatter $\sigma_\text{WL}=0.22$. For each selected cluster, we thus draw a realization of the WL mass. It is then used to compute the mean shear profile based on the \textit{Rubin}-LSST-like source redshift distribution following our extraction model, Eq.~\ref{eq:extraction_model} with no cluster member contamination, and under the assumption of BCG centres. This profile is perturbed by a realization of the shape noise and uncorrelated LSS noise, whose covariance is determined according to the method outlined in Sect.~\ref{sec:extraction_model}.

\begin{figure}[]
    \centering
    \includegraphics[width=\columnwidth]{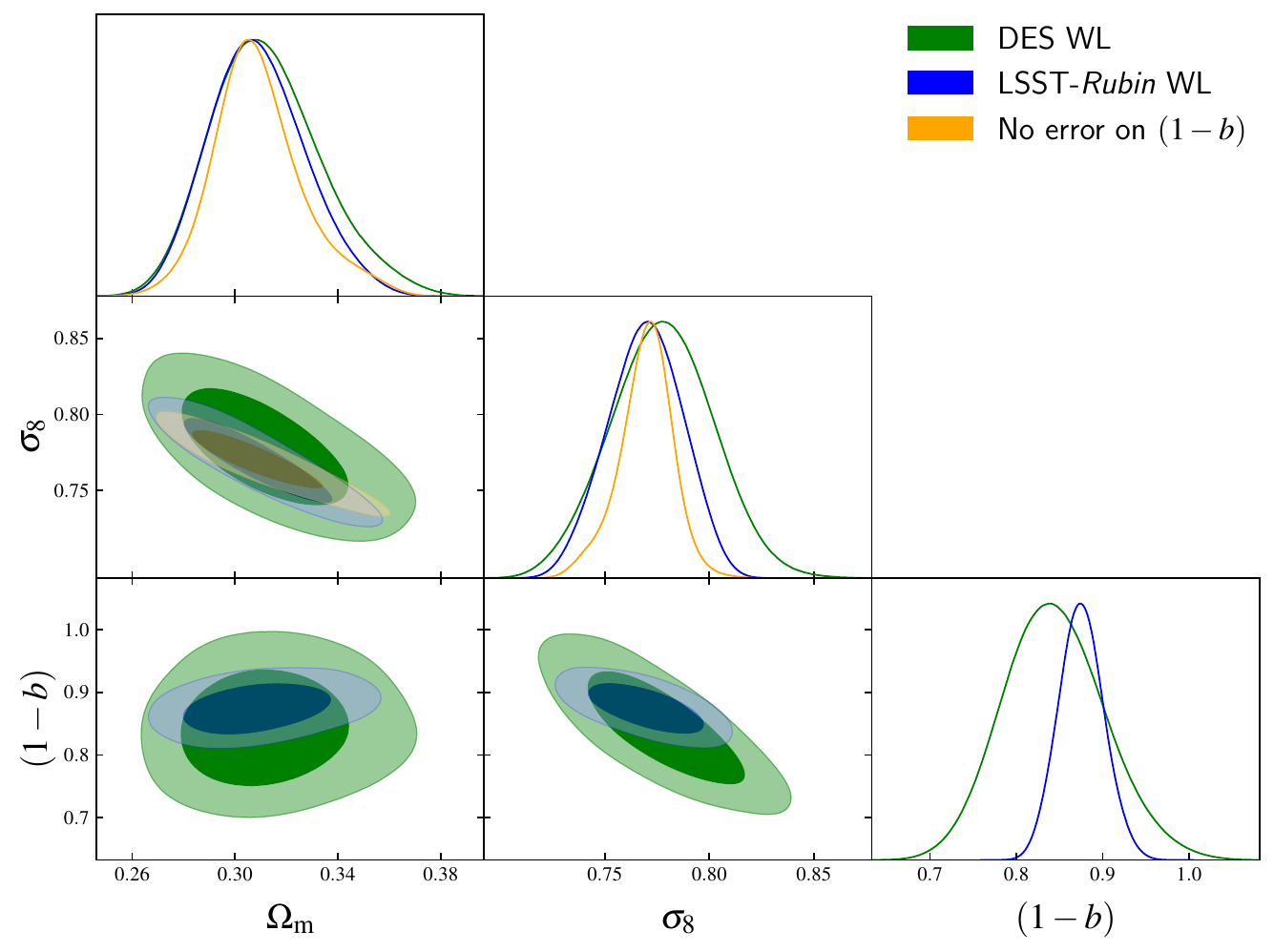}
    \caption{Constraining power forecast for LSST-\textit{Rubin}-like WL data, compared to the current baseline analysis with DES WL data, and no error on the mass bias.}
    \label{fig:future_constraints}
\end{figure}

Using the mock shear profiles instead of those of the DES calibration sample, we constrain cosmological parameters and the mass bias following the process described in Sect.~\ref{constraints}. We also sample the number counts with a perfectly known mass bias to evaluate the maximum statistical constraining potential of the PSZ2 cosmological sample. Figure~\ref{fig:future_constraints} presents the constraints for the mock LSST-\textit{Rubin}-like dataset and the fixed hydrostatic mass bias as well as the baseline analysis with the DES data and \cite{planckcollaborationvi_planck_2020} constraints for comparison.

The constraining power is strongly increased with the mock Stage IV lensing data, particularly in the $S_8$ direction. Interestingly, the constraints are not improved much by completely removing the error on the mass bias (essentially identical constraining power on $S_8$, $\sim$10\% increase on $\Omega_\mathrm{m}$ and $\sim$25\% increase on $\sigma_8$). This shows that while the constraining power of the PSZ2 cluster sample can be greatly improved with Stage IV WL, it will be close to statistics limited, even though we assumed the same mis-centring distribution as the current analysis. We note that this can only be achieved if the uncertainty on the WL mass bias is reduced from $2-4\%$ in the current analysis to $1\%$, which will require a more extensive study of WL systematics.

\section{Conclusion}
\label{conclusion}
We introduced a method to simultaneously calibrate a mean hydrostatic mass bias and constrain the cosmology with a population of clusters with hydrostatic masses computed from ICM observations and WL shear profiles from wide-field galaxy surveys. We applied this method on the \textit{Planck} PSZ2 cosmological sample, with hydrostatic masses derived using a scaling relation calibrated with \textit{Chandra} follow-up X-ray observations, and shear profiles derived from the DES Y3 shape catalogue.
We used three different centre definitions for the derivation of shear profiles (peak of SZ emission, automatic optical counterpart definition, and visual identification of BCG) and calibrated the mis-centring distributions with X-ray derived centres to take them into account in the mass calibration. We then sampled the cosmology and mass calibration simultaneously for a variety of different analysis settings, obtaining $\Omega_\text{m}=0.312^{+0.018}_{-0.024}$, $\sigma_8=0.777\pm 0.024$, $S_8=0.791^{+0.023}_{-0.021}$, and $(1-b)=0.844^{+0.055}_{-0.062}$ for our baseline settings. When considering a hydrostatic mass bias evolving with mass (as favoured by the data), we find $\Omega_\text{m}=0.353^{+0.025}_{-0.031}$, $\sigma_8=0.751\pm 0.023$, $S_8=0.814^{+0.019}_{-0.020}$, $(1-b)=0.850^{+0.055}_{-0.062}$, and $\alpha=0.272\pm0.094$. Compared to other cosmological analyses, these results are in line with previous studies of \textit{Planck} and SPT cluster samples, and more generally the majority of analyses of late-time probes, finding a lower $S_8$ value than \cite{planckcollaborationvi_planck_2020}. Even though we used the same WL data for the mass calibration, our results are in 2.9$\sigma$ tension with the constraints derived from the eRASS1 cluster sample in \cite{ghirardini_srg_2024}, which found a higher $S_8$ value than \cite{planckcollaborationvi_planck_2020}. We verified that the masses we obtained from our best-fit model are compatible with those derived from the best-fit model of \cite{ghirardini_srg_2024}, confirming that the difference in cosmological constraints does not come from the mass calibration. The value of the hydrostatic mass bias that we obtained is in line with other results from both simulations and observations. For comparisons with other observations, instrumental calibration effects need to be taken into account, and help explain the slightly higher $(1-b)$ value we derived. Lastly, we provided a forecast of the possible future constraining power of the PSZ2 cosmological sample with shear profiles from Stage IV WL surveys, showing that the statistical power of the sample is sufficient to obtain much tighter constraints with a more accurate mass calibration.

\section*{Data availability}
The MCMC chains for the BCG 800 $\textsc{bpz}$ and BCG 800 $\textsc{bpz}$ $\alpha$ analyses are available at: \url{https://szdb.osups.universite-paris-saclay.fr/cosmo.html}.

\begin{acknowledgements}
    The authors thank Nabila Aghanim for the fruitful discussions and comments.
  GA acknowledges financial support from the AMX program. 
  This work was supported by the French Space Agency (CNES). 
  This research has made use of the computation facility of the Integrated Data and Operation Center (IDOC, \url{https://idoc.ias.u-psud.fr}) at the Institut d’Astrophysique Spatiale (IAS), 
  as well as the SZ-Cluster Database (\url{https://szdb.osups.universite-paris-saclay.fr}).
\\
  Funding for the DES Projects has been provided by the U.S. Department of Energy, the U.S. National Science Foundation, the Ministry of Science and Education of Spain, 
the Science and Technology Facilities Council of the United Kingdom, the Higher Education Funding Council for England, the National Center for Supercomputing 
Applications at the University of Illinois at Urbana-Champaign, the Kavli Institute of Cosmological Physics at the University of Chicago, 
the Center for Cosmology and Astro-Particle Physics at the Ohio State University,
the Mitchell Institute for Fundamental Physics and Astronomy at Texas A\&M University, Financiadora de Estudos e Projetos, 
Funda{\c c}{\~a}o Carlos Chagas Filho de Amparo {\`a} Pesquisa do Estado do Rio de Janeiro, Conselho Nacional de Desenvolvimento Cient{\'i}fico e Tecnol{\'o}gico and 
the Minist{\'e}rio da Ci{\^e}ncia, Tecnologia e Inova{\c c}{\~a}o, the Deutsche Forschungsgemeinschaft and the Collaborating Institutions in the Dark Energy Survey. 
\\
The Collaborating Institutions are Argonne National Laboratory, the University of California at Santa Cruz, the University of Cambridge, Centro de Investigaciones Energ{\'e}ticas, 
Medioambientales y Tecnol{\'o}gicas-Madrid, the University of Chicago, University College London, the DES-Brazil Consortium, the University of Edinburgh, 
the Eidgen{\"o}ssische Technische Hochschule (ETH) Z{\"u}rich, 
Fermi National Accelerator Laboratory, the University of Illinois at Urbana-Champaign, the Institut de Ci{\`e}ncies de l'Espai (IEEC/CSIC), 
the Institut de F{\'i}sica d'Altes Energies, Lawrence Berkeley National Laboratory, the Ludwig-Maximilians Universit{\"a}t M{\"u}nchen and the associated Excellence Cluster Universe, 
the University of Michigan, NSF NOIRLab, the University of Nottingham, The Ohio State University, the University of Pennsylvania, the University of Portsmouth, 
SLAC National Accelerator Laboratory, Stanford University, the University of Sussex, Texas A\&M University, and the OzDES Membership Consortium.
\\
Based in part on observations at NSF Cerro Tololo Inter-American Observatory at NSF NOIRLab (NOIRLab Prop. ID 2012B-0001; PI: J. Frieman), which is managed by the Association of Universities for Research in Astronomy (AURA) under a cooperative agreement with the National Science Foundation.
\\
The DES data management system is supported by the National Science Foundation under Grant Numbers AST-1138766 and AST-1536171.
The DES participants from Spanish institutions are partially supported by MICINN under grants PID2021-123012, PID2021-128989 PID2022-141079, SEV-2016-0588, CEX2020-001058-M and CEX2020-001007-S, some of which include ERDF funds from the European Union. IFAE is partially funded by the CERCA program of the Generalitat de Catalunya.
\\
We  acknowledge support from the Brazilian Instituto Nacional de Ci\^encia
e Tecnologia (INCT) do e-Universo (CNPq grant 465376/2014-2).
\\
This document was prepared by the DES Collaboration using the resources of the Fermi National Accelerator Laboratory (Fermilab), a U.S. Department of Energy, Office of Science, Office of High Energy Physics HEP User Facility. Fermilab is managed by Fermi Forward Discovery Group, LLC, acting under Contract No. 89243024CSC000002.
\end{acknowledgements}

\section*{Author Contribution}

GA performed the main analysis, produced the majority of the plots in the paper, and led the paper writing. SG measured and calibrated the DES WL signal around the Planck clusters, and contributed to the visual identification of BCG, to the inclusion of the DES WL signal in the modelling pipeline, and to the writing and plot making. MD, GWP, and LS provided direct supervision for the research and contributed to the paper writing. FAS, WRF, and CJ contributed the Chandra data. SB and MC acted as DES internal reviewers. The remaining authors have made contributions to this paper that include, but are not limited to, the construction of DECam and other aspects of collecting the data; data processing and calibration; developing broadly used methods, codes, and simulations; running the pipelines and validation tests; and promoting the science analysis.

\bibliography{biblio}
\bibliographystyle{aa}

\begin{appendix}

\section{Cluster Member Contamination}
\label{cluster_member_contamination}
\begin{figure*}[]
    \centering
    \includegraphics[width=\textwidth]{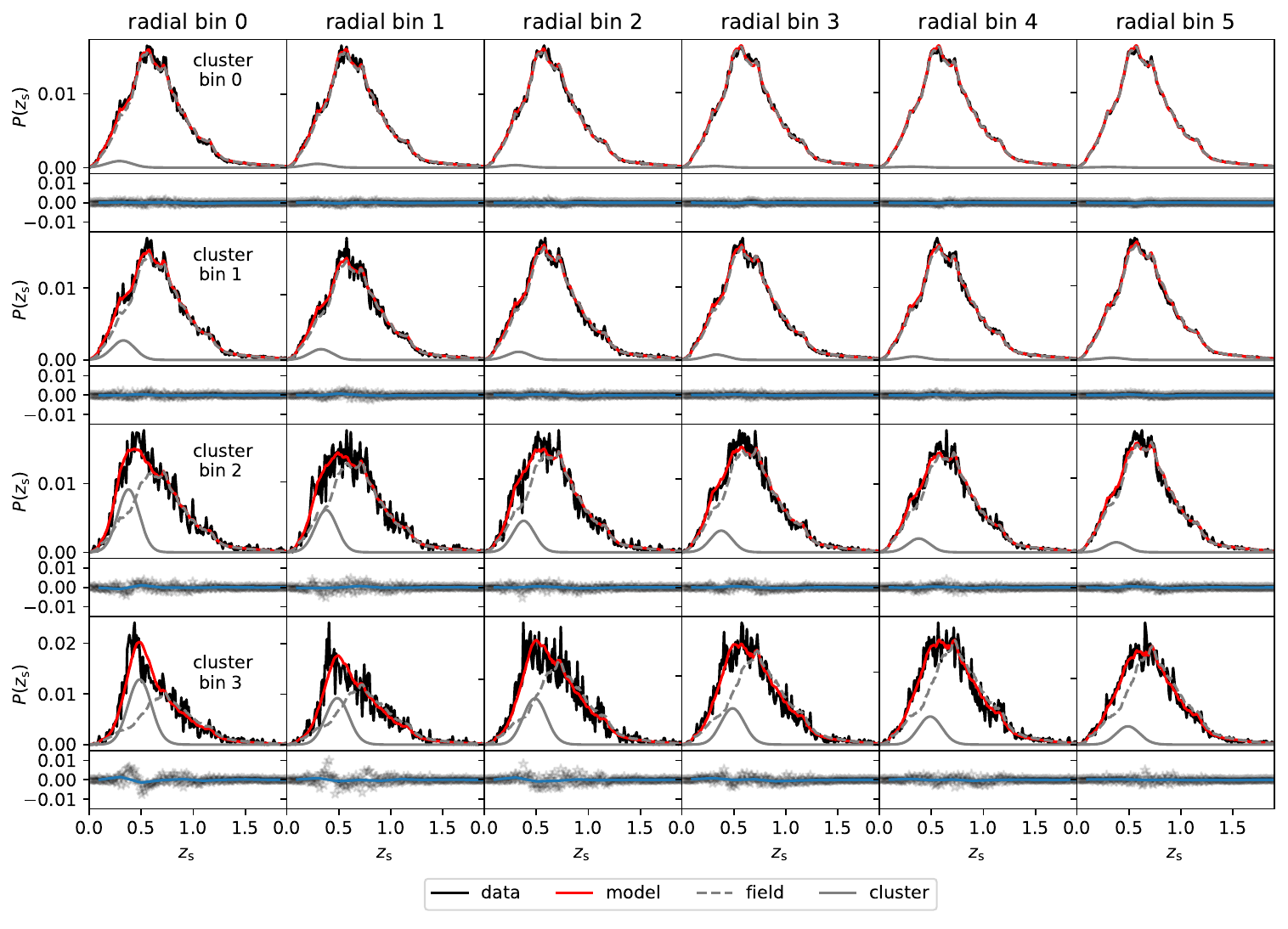}
    \caption{Source redshift decomposition. For each cluster bin and radial bin, we show in black the stacked empirical source redshift distribution, and in red the best-fit model. The residuals between the two are shown in the smaller inserts below each panel. The model is the sum of the cluster component (full gray line) and the field component (dashed gray line). The blue line shows the mean of the residuals, binned in coarse source redshift bins.}
    \label{fig:cluster_member_decomp}
\end{figure*}
Galaxy clusters appear in photometric data as overdensities of galaxies. As such, even in the presence of a background selection for WL sources, a noticeable amount of galaxies associated with the clusters is bound to leak into the selected background sample. These galaxies are not lensed by the cluster potential. They thus bias the measured shear profile low w.r.t. the true shear profile. This problem is called cluster member contamination.

The degree of cluster member contamination can be assessed empirically. For once, the presence of cluster members in the background selection leads to an increase in the measured source number density towards the cluster centre. However, the source number density is also impacted by magnification and blending. Correctly extracting the cluster member contamination signal from the source number density profiles thus requires careful modelling and image injections \citep{kleinebreil_srg_2025}.

Instead, we determine the cluster member contamination from the radial trends in the measured source redshift distribution, as first proposed by \citet{gruen_weak_2014}. We follow closely the approach of \citet{bocquet_spt_2024, grandis_srg_2024}. In short, the radial trend of the source redshift distribution is modelled by two components,
\begin{equation}
    P(z_\text{s} | R, \lambda, z_\text{cl}) = f_\text{cl}(R | \lambda, z_\text{cl}) P_\text{cl}(z_\text{s} | z) + \left[1- f_\text{cl}(R | \lambda, z_\text{cl}) \right] P_\text{fld}(z_\text{s}),
\end{equation}
with 
\begin{equation}
    P_\text{cl}(z_\text{s} | z) = \mathcal{N} \left( z_\text{s}| \mu_\text{clm}(z_\text{cl}), \sigma^2_\text{clm}(z_\text{cl}) \right)
\end{equation}
The first component describes the cluster members with a Gaussian with mean $\mu_\text{clm}(z_\text{cl})$ and variance $\sigma^2_\text{clm}(z_\text{cl})$ that are fitted from the data as linear functions of cluster redshift $z_\text{cl}$. The second component describes the local background galaxies, which are expected to follow the same redshift distribution as the distribution in the field, $P_\text{fld}(z_\text{s})$. Finally, the radial mixture $f_\text{cl}(R | \lambda, z_\text{cl})$ has a physical parametrization based on an NFW profile, the cluster richness $\lambda$, and the distance from the cluster $R$ \citep[see][Appendix B for an in-depth explanation]{grandis_srg_2024}. All clusters in the sample are simultaneously co-fitted with this model, as described in \citet{grandis_srg_2024}, section 3.3.1. 

We perform this analysis once for $\textsc{bpz}$ and once for $\textsc{dnf}$ source redshift estimates. For the four cluster bins defined in Sect.~\ref{gof}, we show the stacked source redshift distributions defined in black, as well as the best-fit model in orange, the cluster components in gray, and the downscaled field component in dashed gray for different radial bins in Fig.~\ref{fig:cluster_member_decomp}. The increase of the cluster component is evident by the change in the shape of the distribution towards the cluster centre. The model provides a good fit to the data, as can be seen by the residuals shown below each panel.

\begin{figure}[]
    \centering
    \includegraphics[width=\columnwidth]{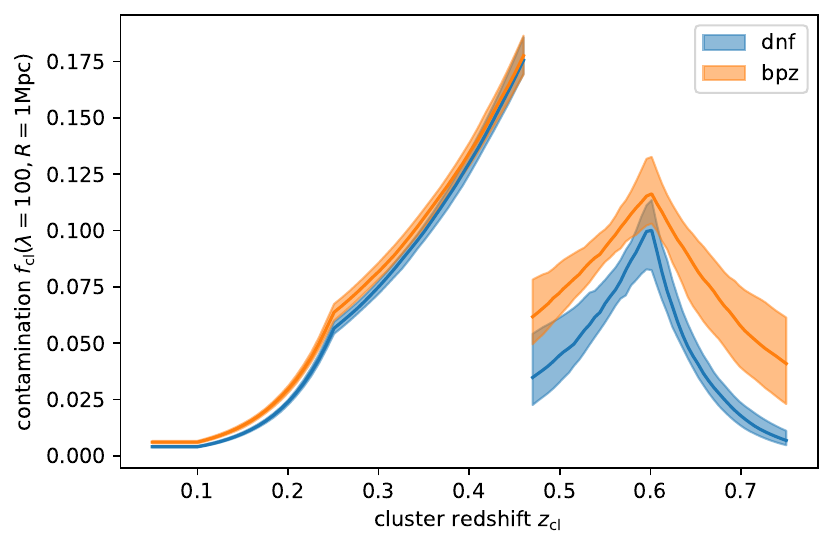}
    \caption{Cluster member contamination for a richness $\lambda=100$ cluster at a cluster centric distance $R=1$ Mpc as a function of cluster redshift $z_\text{cl}$, as fitted from the empirical source redshift distribution of $\textsc{dnf}$ and $\textsc{bpz}$ photometric redshift estimates for the DES WL sources.}
    \label{fig:fcl}
\end{figure}
In this work, we undertake one improvement compared to the analysis in \citet{grandis_srg_2024}. Both works stop using the second tomographic redshift bin for clusters whose redshift is larger than the median redshift of the bin, $z_\text{cl}>z_{\text{med},b=2}=0.47$. Given the width of the source redshift distribution of the tomographic bins, many cluster members will be included in our source sample for clusters with $z_\text{cl}< z_{\text{med},b=2}$, as their members naturally fall in the tomographic redshift bin 2.  Once that bin is excluded for $z_\text{cl}>z_{\text{med},b=2}$, only a small fraction of cluster members will leak into the following tomographic bins as the latter are at much higher redshift. It is therefore plausible to expect the contamination by cluster members as a function of cluster redshift to abruptly diminish at the median redshifts of the tomographic redshift bins \citep{singh_galaxy_2025}. This effect results from the background selection we used (see eq.~\ref{eq:tomo_weights}). 

Compared to \citet{grandis_srg_2024}, we modify the model to allow for a discontinuity in the cluster member contamination at $z_\text{cl}=0.47$, the median redshift of the second tomographic redshift bin of DES WL sources, and the cluster redshift at which we drop that tomographic bin. As shown in Fig.~\ref{fig:fcl}, our best-fit model recovers a noticeable discontinuity at that redshift.

As the cluster member contamination recovered from the $\textsc{bpz}$ redshift estimates is larger than that recovered from the $\textsc{dnf}$, we perform two cosmological analyses, using the two different best-fits. As shown in Sect.~\ref{constraints}, the resulting cosmological parameters are very similar. This demonstrates that our handling of cluster member contamination is stable.

  \section{WL mass bias from synthetic shear profiles}
  \label{synth_shear_profiles}
  The fact that we use an imperfect shear extraction model, the inherent heterogeneity of cluster mass profiles, and our residual systematic uncertainty on specific aspects of the WL data, like the multiplicative shear bias, require that the mapping between the WL mass $M_\text{WL}$ and the halo mass be calibrated with dedicated simulations \citep{grandis_calibration_2021, bocquet_spt_2024, grandis_srg_2024}.

  We created these simulations for the different observational settings we adopted. Our baseline is to use BCG centres to limit our analysis to scale $R_\text{min}>800h^{-1} $Mpc and to use the photometric redshifts from $\textsc{bpz}$ to estimate the cluster member contamination. Each of these modeling choices is altered one by one to explore the sensitivity of our results. We also consider MCMF centres and SZ centres, consider smaller scales $R_\text{min}>500h^{-1} $Mpc, and use photometric redshifts from $\textsc{dnf}$ to estimate the cluster member contamination. In all, this leads to 5 settings that require a WL mass bias determination: BCG 800 \textsc{bpz}, BCG 500 \textsc{bpz}, BCG 800 \textsc{dnf}, SZ 800 \textsc{bpz}, MCMF 800 \textsc{bpz}.

  For each setting, we create $\sim 1k$ Monte Carlo simulations, varying the source redshift distribution and multiplicative shear bias within the prior derived by \citet{myles_dark_2021} and used in the main DES Y3 cosmic shear and galaxy correlation study \citep{amon_dark_2022}. We vary the mis-centring distribution of the specific centre within the priors derived in this work. The chosen cluster member contamination also varies with the prior derived above.

  In each realization, we draw a synthetic shear profile for a sample of halos and surface mass densities taken from the cosmological hydrodynamical TNG300 simulations \citep{pillepich_simulating_2018, springel_first_2018} in four redshift snapshots, $z_\text{snap}=0.11,\,0.24,\,0.42,\,0.64$. Each selected halo is assigned 
  \begin{enumerate}
      \item the redshift of its snapshot,
      \item the halo mass $M_\text{500c}$ based on the gravity only run of the simulation,
      \item a synthetic source redshift distribution of DES-Y3 source, reflective of our background selection (see Eq.~\ref{eq:tomo_weights}),
      \item a richness drawn from the richness mass relation calibrated by \citet{chiu_erosita_2022} to assign a cluster member contamination profile based on a draw from the best-fit derived in Sect.~\ref{cluster_member_contamination},
      \item a synthetic shear profile considering the azimuthal anisotropies of the projected mass distribution and the source redshift-dependent effect of reduced shear, as derived in Eq. 33 in \citet{grandis_srg_2024}, which is diluted in the cluster member contamination profile to create a noise-free, synthetic reduced shear profile, evaluated for different mis-centring radii,
      \item a best-fit extraction mass $M_\text{WL}$ for each assumed mis-centring resulting from the fit of our shear extraction model to the synthetic shear profiles.
  \end{enumerate}

    For each redshift snapshot $l$ we can thus simulate a WL mass for a range of mis-centring radii. When fitting the WL mass to halo mass relation, we weight each mis-centring radius by the mis-centring distribution, to extract, for each realisation, a set of bias and scatter parameters

    \begin{equation}\label{eq:WLbias}
        \bigg< \ln \frac{M_\text{WL}}{M_0} \bigg| M, z_\text{cl} \bigg> =b^\text{WL}_l + b_\text{M}^\text{WL} \ln \left( \frac{M}{M_0} \right) \text{ and }
\ln \sigma_\text{WL}^2 = s_l,
    \end{equation}
with the mass trend co-fitted over all snapshots.

By repeating the procedure $\sim 1k$ times while varying the known systematics of WL, we determine the WL bias and its systematic uncertainty as a function of redshift, as well as its mass trend with systematic uncertainty, and the WL scatter as a function of redshift, with systematic uncertainty. The WL bias posterior is compressed into a mean value $\mu_{\text{b},l}$ and two principal components $\delta_{\text{b1},l}$ and $\delta_{\text{b2},l}$, scaled by their variances. Realizations of the redshift-dependent WL mass bias can then be obtained by drawing two 
random standard normal variates $A_\text{WL}$, $B_\text{WL}$, and computing
\begin{equation}\label{eq:bWL_w_params}
    b(z_\text{cl}) = \mathcal{I}(z_\text{cl}| z_l,\,\mu_{\text{b},l}) + A_\text{WL}\mathcal{I}(z_\text{cl}| z_l,\,\delta_{\text{b1},l}) + B_\text{WL}\mathcal{I}(z_\text{cl}| z_l,\,\delta_{\text{b2},l}),
\end{equation}
where $\mathcal{I}(z_\text{cl}| z_l,\,\mu_{\text{b},l})$ is the interpolation of the data vectors $( z_l,\,\mu_{\text{b},l})$ 
at redshift $z_\text{cl}$.  Similarly, for the mass trend $b_\text{M} =  \mu_{b_\text{M}} + C_\text{WL} \cdot \sigma_{b_\text{M}}$ we calibrate the mean and standard deviation, and introduce a random standard normal variate $C_\text{WL}$ to sample its systematic uncertainty.

The scatter is similarly modeled as
\begin{equation}\label{eq:sigmaWL}
    \sigma_\text{WL}^2 = \exp \left\{ \mathcal{I}(z_\text{cl}| z_l,\,\mu_{\text{s},l}) + D_\text{WL}\mathcal{I}(z_\text{cl}| z_l,\,\delta_{\text{s},l}) \right\},
\end{equation}
where $\mu_{\text{s},l}$ and $\delta_{\text{s},l}$ are the mean and standard deviations of the WL scatter in the simulation snapshot $l$. Tables B1-5 report the calibration values for the five different settings we considered.

The WL bias and scatter numbers based on BCG centres display no deviations from the previous results for SPT or eROSITA selected clusters \citep{chiu_erosita_2022, bocquet_spt_2024, grandis_srg_2024, kleinebreil_srg_2025}: the bias is almost 0, with a mass trend of unity, and a scatter $\sigma_\text{WL}\approx 0.2$. The situation, however, changes when using SZ centres. In this case, the WL mass bias increases from zero at $z=0.11$ to $20\%$ at $z=0.64$. At the latter redshift, the fixed angular scale of the SZ mis-centring subtends a sizeable physical distance, leading to clear discrepancies between our extraction model and the synthetic data. The choice of MCMF centres leads to a significant increase in the WL scatter to $\sigma_\text{WL}\approx 0.36$ because a small fraction of the clusters are drastically mis-centred, while the extraction model ignores these outliers to provide a good mean model at a given halo mass. Ultimately, these alterations are spot on, as shown by the stability of our inference to the centring choice (see Sect.~\ref{constraints}). 

\begin{table}
\caption{\label{tab:bWL--des}
Calibration of WL bias and scatter for DES Y3 WL on \textit{Planck} clusters considering BCG centres, a minimal fit radius $R_\text{min}>500h^{-1} $Mpc, and $\textsc{bpz}$ based cluster member contamination.}
\begin{tabular}{lcccc}
$z$ & $0.11$& $0.24$& $0.42$& $0.64$\\  
\hline
$\mu_\text{b}$ & $-0.020$ & $-0.015$ & $-0.022$ & $-0.020$ \\  
$\delta_\text{b1}$ & $-0.020$ & $-0.023$ & $-0.028$ & $-0.038$ \\  
$\delta_\text{b2}$ & $0.012$ & $0.010$ & $0.004$ & $-0.016$ \\  
$\mu_\text{b}$ & $-0.020$ & $-0.015$ & $-0.022$ & $-0.020$ \\  
$\mu_\text{s}$ & $-3.343$ & $-3.389$ & $-3.389$ & $-3.282$ \\  
$\delta_\text{s}$ & $0.278$ & $0.280$ & $0.287$ & $0.288$ \\ 
\hline
 &  \multicolumn{4}{c}{$b_\text{M}=1.005 \pm 0.019$} \\ 
\multicolumn{5}{p{\linewidth-12pt} }{\small Numerical values for the WL bias ($\mu_\text{b}$, $\delta_\text{b1}$, $\delta_\text{b2}$) and scatter ($\mu_\text{s}$, $\delta_{s}$) calibration at different redshift $z$, and their global mass trends $b_\text{M}$, and $s_\text{M}$, respectively. }
\end{tabular}
\end{table}

\begin{table}
\caption{Calibration of WL bias and scatter for DES Y3 WL on \textit{Planck} clusters considering BCG centres, a minimal fit radius $R_\text{min}>800h^{-1} $Mpc, and $\textsc{bpz}$ based cluster member contamination.}
\begin{tabular}{lcccc}
$z$ & $0.11$& $0.24$& $0.42$& $0.64$\\  
\hline
$\mu_\text{b}$ & $-0.013$ & $-0.010$ & $-0.024$ & $-0.029$ \\  
$\delta_\text{b1}$ & $-0.020$ & $-0.022$ & $-0.027$ & $-0.037$ \\  
$\delta_\text{b2}$ & $0.010$ & $0.009$ & $0.005$ & $-0.014$ \\  
$\mu_\text{b}$ & $-0.013$ & $-0.010$ & $-0.024$ & $-0.029$ \\  
$\mu_\text{s}$ & $-3.285$ & $-3.311$ & $-3.282$ & $-3.171$ \\  
$\delta_\text{s}$ & $0.257$ & $0.261$ & $0.261$ & $0.265$ \\  
\hline
 &  \multicolumn{4}{c}{$b_\text{M}=1.018 \pm 0.019$} \\
 \multicolumn{5}{p{\linewidth-12pt} }{\small Same as Table~\ref{tab:bWL--des} }
\end{tabular}
\end{table}

\begin{table}
\caption{Calibration of WL bias and scatter for DES Y3 WL on \textit{Planck} clusters considering BCG centres, a minimal fit radius $R_\text{min}>800h^{-1} $Mpc, and $\textsc{dnf}$ based cluster member contamination.}
\begin{tabular}{lcccc}
$z$ & $0.11$& $0.24$& $0.42$& $0.64$\\  
\hline
$\mu_\text{b}$ & $-0.013$ & $-0.010$ & $-0.025$ & $-0.028$ \\  
$\delta_\text{b1}$ & $-0.019$ & $-0.022$ & $-0.026$ & $-0.036$ \\  
$\delta_\text{b2}$ & $0.011$ & $0.009$ & $0.005$ & $-0.015$ \\  
$\mu_\text{b}$ & $-0.013$ & $-0.010$ & $-0.025$ & $-0.028$ \\  
$\mu_\text{s}$ & $-3.287$ & $-3.312$ & $-3.285$ & $-3.173$ \\  
$\delta_\text{s}$ & $0.262$ & $0.263$ & $0.265$ & $0.265$ \\  
\hline
 &  \multicolumn{4}{c}{$b_\text{M}=1.018 \pm 0.019$}  \\
 \multicolumn{5}{p{\linewidth-12pt} }{\small Same as Table~\ref{tab:bWL--des} }
\end{tabular}
\end{table}

\begin{table}
\caption{Calibration of WL bias and scatter for DES Y3 WL on \textit{Planck} clusters considering MCMF centres, a minimal fit radius $R_\text{min}>800h^{-1} $Mpc, and $\textsc{bpz}$ based cluster member contamination.}
\begin{tabular}{lcccc}
$z$ & $0.11$& $0.24$& $0.42$& $0.64$\\  
\hline
$\mu_\text{b}$ & $-0.021$ & $-0.004$ & $-0.005$ & $-0.000$ \\  
$\delta_\text{b1}$ & $0.039$ & $0.029$ & $0.031$ & $0.041$ \\  
$\delta_\text{b2}$ & $0.041$ & $-0.012$ & $-0.011$ & $-0.022$ \\  
$\mu_\text{b}$ & $-0.021$ & $-0.004$ & $-0.005$ & $-0.000$ \\  
$\mu_\text{s}$ & $-1.993$ & $-2.022$ & $-2.088$ & $-2.008$ \\  
$\delta_\text{s}$ & $0.427$ & $0.411$ & $0.369$ & $0.334$ \\  
\hline
 &  \multicolumn{4}{c}{$b_\text{M}=0.980 \pm 0.021$} \\
 \multicolumn{5}{p{\linewidth-12pt} }{\small Same as Table~\ref{tab:bWL--des} }
\end{tabular}
\end{table}

\begin{table}
\caption{Calibration of WL bias and scatter for DES Y3 WL on \textit{Planck} clusters considering SZ centres, a minimal fit radius $R_\text{min}>800h^{-1} $Mpc, and $\textsc{bpz}$ based cluster member contamination.}
\begin{tabular}{lcccc}
$z$ & $0.11$& $0.24$& $0.42$& $0.64$\\  
\hline
$\mu_\text{b}$ & $0.001$ & $0.075$ & $0.138$ & $0.200$ \\  
$\delta_\text{b1}$ & $-0.016$ & $-0.018$ & $-0.025$ & $-0.045$ \\  
$\delta_\text{b2}$ & $0.011$ & $0.012$ & $0.017$ & $-0.018$ \\  
$\mu_\text{b}$ & $0.001$ & $0.075$ & $0.138$ & $0.200$ \\  
$\mu_\text{s}$ & $-3.198$ & $-3.245$ & $-3.034$ & $-2.669$ \\  
$\delta_\text{s}$ & $0.255$ & $0.258$ & $0.270$ & $0.295$ \\  
\hline
 &  \multicolumn{4}{c}{$b_\text{M}=0.988 \pm 0.019$}  \\
 \multicolumn{5}{p{\linewidth-12pt} }{\small Same as Table~\ref{tab:bWL--des} }
\end{tabular}
\end{table}

\section{Impact of the BAO likelihood}
\label{BAO_impact}
\begin{figure*}
    \centering
    \resizebox{\hsize}{!}
    {\includegraphics{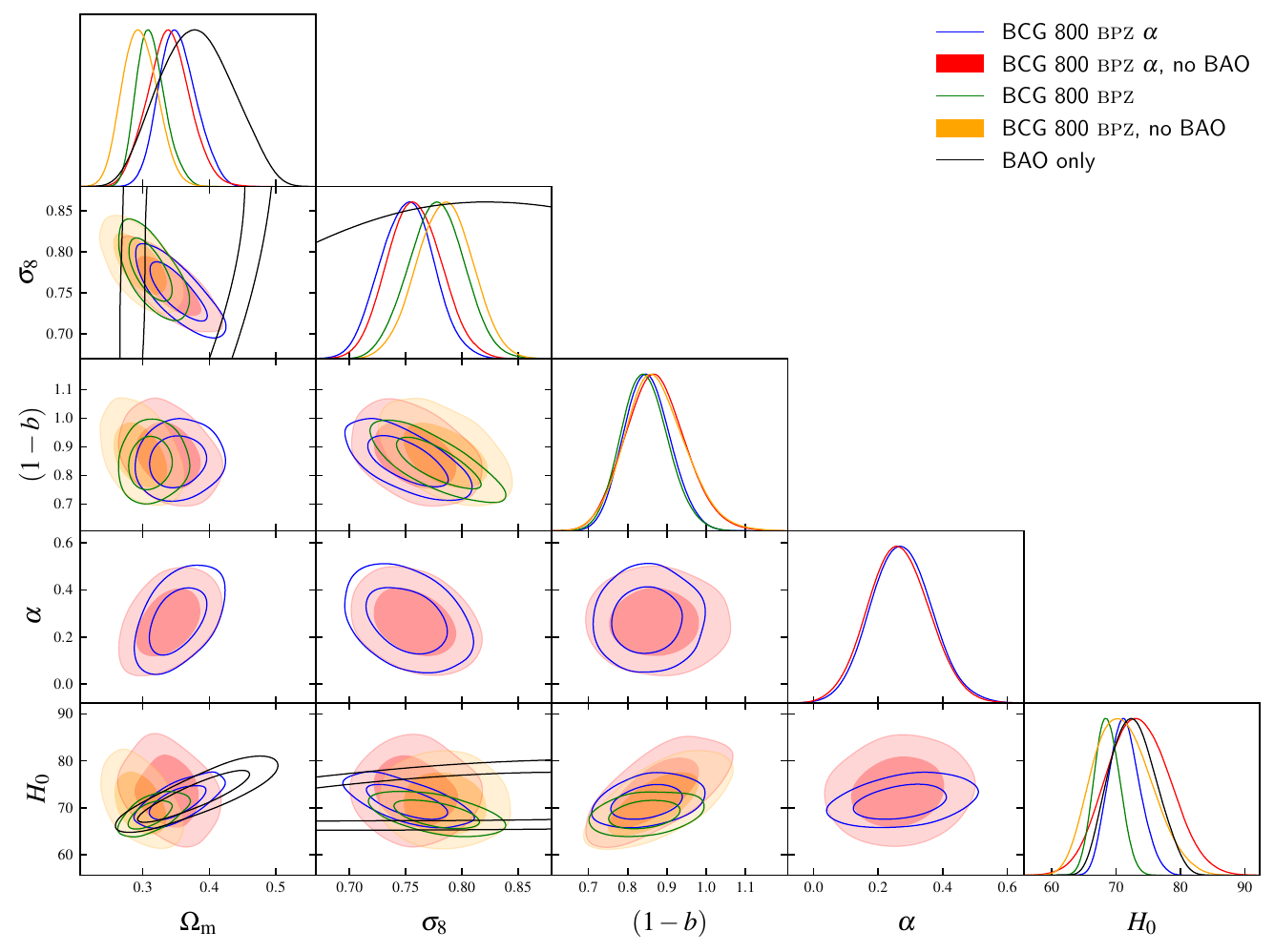}}
    \caption{Comparison of the constraints obtained with and without inclusion of the BAO likelihood, for the BCG 800 \textsc{bpz} and BCG 800 \textsc{bpz} $\alpha$ configurations.}
    \label{fig:noBAO}
\end{figure*}

In this Appendix, we study the impact of the BAO likelihood by comparing the constraints obtained without its inclusion with those presented in the main text. This is particularly interesting in the context of comparing our results with other cluster analyses that do not include BAO data. We choose to use the same prior as in \citet{bocquet_spt_2024a}, that is $H_0 \sim \mathcal{N}(70,5)$. We note that the analysis of eROSITA cluster number counts presented in \citet{ghirardini_srg_2024} uses a much tighter prior taken from \citet{planckcollaborationvi_planck_2020}. Figure~\ref{fig:noBAO} presents the results obtained with and without the BAO likelihood for the BCG 800 \textsc{bpz} and BCG 800 \textsc{bpz} $\alpha$ configurations, as well as the constraints obtained by sampling the BAO likelihood on its own.

We find that removing the BAO likelihood and using the same prior on $H_0$ as \citet{bocquet_spt_2024a} weakens the final constraints slightly, yielding: $\Omega_\text{m} = 0.296^{+0.024}_{-0.028}$, $\sigma_8 = 0.785\pm 0.024$ and $S_8 = 0.777\pm 0.029$. The BAO data imposes a tight correlation between $H_0$ and $\Omega_\mathrm{m}$, and helps to constrain the lower bound of $\Omega_\mathrm{m}$, as shown by the corresponding panel in Fig.~\ref{fig:noBAO}. This, combined with the $\Omega_\mathrm{m}$ constraining of the cluster number counts, provides a tight constraint on $H_0$. Because $H_0$ enters into the distance measurements necessary to the WL likelihood, the inclusion of the BAO likelihood leads to a more precise mass calibration, which impacts the constraining power of the number counts likelihood. 

Overall, the inclusion of the BAO likelihood provides some additional constraining power to the lower bound of $\Omega_\mathrm{m}$ and tightens the $S_8$ constraints slightly by improving the WL mass calibration, but does not significantly change any conclusions of our analysis.

\newpage

\section{Full triangle plot of MCMC fitting}
\begin{figure*}
    \centering
    \resizebox{0.99\hsize}{!}
          {\includegraphics{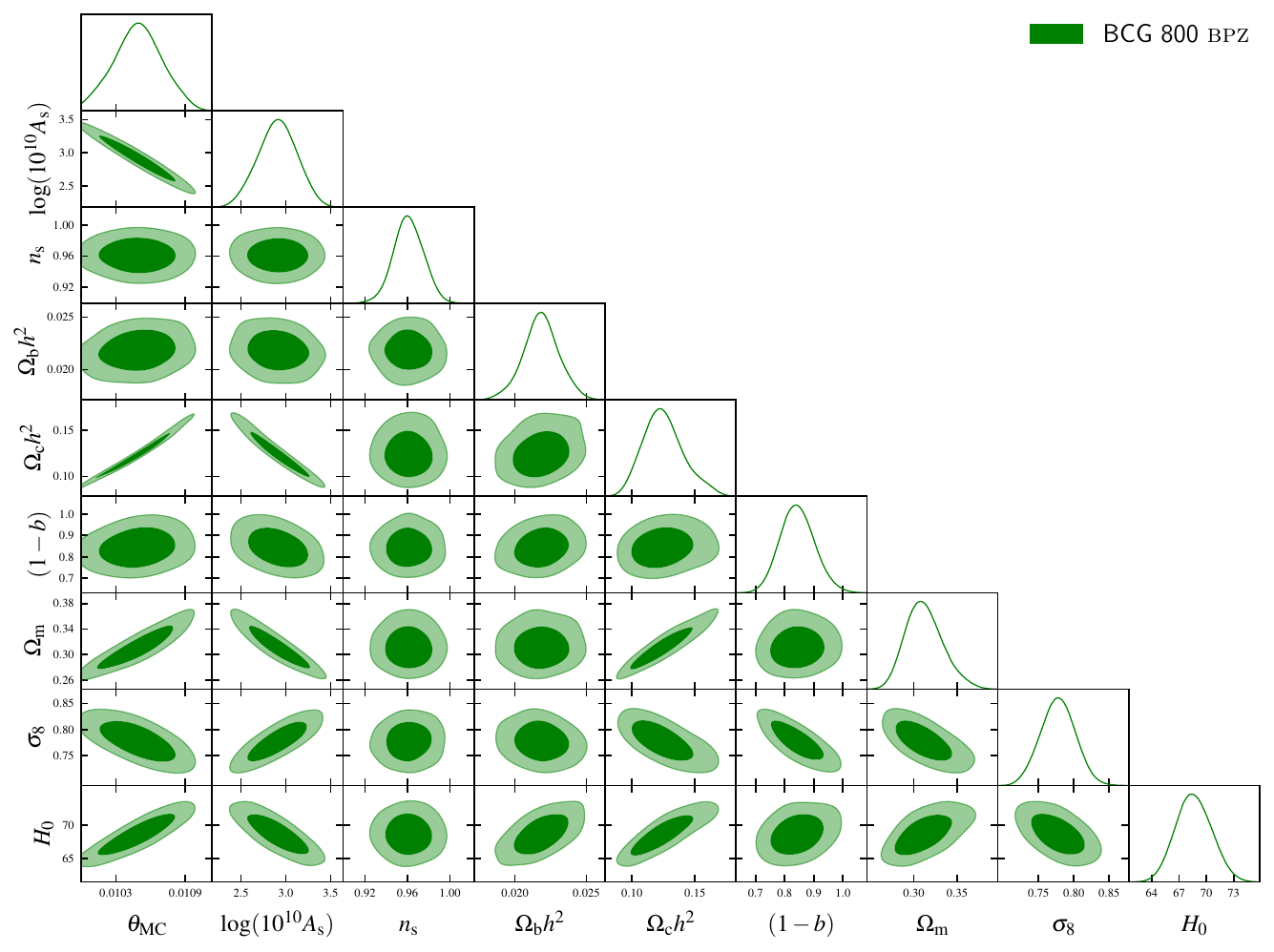}}
    \caption{Parameter constraints obtained from the MCMC fitting of the number counts for the BCG 800 $\textsc{bpz}$ analysis.}
    \label{fig:full_triangle}
\end{figure*}

Fig.~\ref{fig:full_triangle} shows the full parameter constraints of the MCMC sampling of the number counts and mass calibration likelihood.
\end{appendix}

\end{document}